\documentclass[manuscript]{acmart}

\usepackage[
type={CC},
modifier={by-nc},
version={4.0},
imagemodifier={-eu-88x31},
]{doclicense}

\usepackage{listings}
\usepackage{amsthm}
\usepackage{xspace}
\usepackage{textcomp}
\usepackage{mathtools}
\usepackage{mathrsfs}
\usepackage{algorithm}
\usepackage{algorithmic}
\usepackage{threeparttable}
\usepackage{array}
\newcommand{\PreserveBackslash}[1]{\let\temp=\\#1\let\\=\temp}
\newcolumntype{C}[1]{>{\PreserveBackslash\centering}p{#1}}
\newcolumntype{R}[1]{>{\PreserveBackslash\raggedleft}p{#1}}
\newcolumntype{L}[1]{>{\PreserveBackslash\raggedright}p{#1}}
\usepackage{subfigure}
\usepackage{url}
\usepackage{graphicx}
\usepackage{booktabs}
\usepackage{multirow}
\usepackage{dcolumn}
\newlength{\thinline}
\newlength{\thickline}
\usepackage{listings}
\usepackage{caption}
\usepackage{amsmath}
\usepackage{booktabs}
\usepackage[T1]{fontenc}
\usepackage{enumitem}
\setenumerate[1]{itemsep=0pt,partopsep=0pt,parsep=\parskip,topsep=5pt}
\setitemize[1]{itemsep=0pt,partopsep=0pt,parsep=\parskip,topsep=5pt}
\setdescription{itemsep=0pt,partopsep=0pt,parsep=\parskip,topsep=5pt}
\usepackage{textcomp}
\usepackage{booktabs}
\usepackage{multirow}
\usepackage{dcolumn}
\usepackage{amsmath}
\usepackage{threeparttable}
\usepackage{colortbl}
\definecolor{mygray}{gray}{0.9}
\definecolor{deepgray}{gray}{.8}
\definecolor{myblue}{cmyk}{.3,0,0,0}
\definecolor{deepblue}{cmyk}{.3,0,0,0}
\usepackage[utf8]{inputenc}
\usepackage[english]{babel}
\usepackage{amsthm}
  
\usepackage[algo2e]{algorithm2e}
\usepackage{algorithm}
\usepackage{array}
\usepackage{verbatim}
\usepackage[inkscapelatex=false]{svg}
\usepackage{fancyvrb}
\usepackage[most]{tcolorbox}

\newcommand{\name}{\textsc{ThreatPilot}\xspace}
\usepackage{tabularx}
\usepackage{varwidth}
\usepackage{hyperref}

\definecolor{SoftRed}{RGB}{255,200,200}
\definecolor{SoftGreen}{RGB}{230, 238, 213}
\definecolor{SoftPurple}{RGB}{220,200,255}
\definecolor{SoftBlue}{RGB}{200,220,255}

\newcommand{\coloredbox}[2]{%
    \colorbox{#1}{\begin{varwidth}[t]{\dimexpr\linewidth-2\fboxsep\relax}#2\end{varwidth}}%
}
\usepackage{pifont}
\usepackage{dblfloatfix}
\usepackage{dblfloatfix}
\usepackage{dblfloatfix}
\usepackage{dblfloatfix}
\usepackage{dblfloatfix}
\usepackage{dblfloatfix}

\definecolor{lightpurple}{RGB}{235, 239, 249}
\definecolor{lightyellow}{RGB}{238, 197, 145}
\definecolor{orange}{RGB}{255, 165, 0}
\definecolor{SoftRed}{RGB}{255, 204, 204}
\definecolor{mydarkgreen}{RGB}{0,100,0}

\setcopyright{none}
\makeatletter
\renewcommand\@copyrightpermission{}
\makeatother

\begin{document}

\title{\name: Attack-Driven Threat Intelligence Extraction} 

\author{Ming Xu}
\affiliation{%
  \institution{National University of Singapore}
  \city{Singapore}
  \country{Singapore}}
\email{mingxucs18@gmail.com}

\author{Hongtai Wang}
\affiliation{%
  \institution{National University of Singapore}
  \city{Singapore}
  \country{Singapore}}
\email{wanghongtai0702@gmail.com}

\author{Jiahao Liu}
\affiliation{%
  \institution{National University of Singapore}
  \city{Singapore}
  \country{Singapore}}
\email{jiahao99@comp.nus.edu.sg}

\author{Xinfeng Li}
\affiliation{%
  \institution{Nanyang Technological University}
  \city{Singapore}
  \country{Singapore}}
\email{lxfmakeit@gmail.com}

\author{Zhengmin Yu}
\affiliation{%
  \institution{Fudan University}
  \city{Shanghai}
  \country{China}}
\email{zmyu23@m.fudan.edu.cn} 

\author{Weili Han}
\affiliation{%
  \institution{Fudan University}
  \city{Shanghai}
  \country{China}}
\email{wlhan@fudan.edu.cn}

\author{Hoon Wei Lim}
\affiliation{%
  \institution{Cyber Special Ops-R\&D, NCS Group}
  \city{Singapore}
  \country{Singapore}}
\email{hoonwei.lim@ncs.com.sg}

\author{Jin Song Dong}
\affiliation{%
  \institution{National University of Singapore}
  \city{Singapore}
  \country{Singapore}}
\email{dcsdjs@nus.edu.sg}

\author{Jiaheng Zhang}
\affiliation{%
  \institution{National University of Singapore}
  \city{Singapore}
  \country{Singapore}}
\email{jhzhang@nus.edu.sg}

\begin{abstract}
Efficient defense against dynamically evolving advanced persistent threats (APT) requires the structured threat intelligence feeds, such as techniques used.   
However, existing threat-intelligence extraction techniques predominantly focus on individual pieces of intelligence--such as isolated techniques or atomic indicators--resulting in fragmented and incomplete representations of real-world attacks. This granularity inherently limits on both the depth and the contextual richness of the extracted intelligence, making it difficult for downstream security systems to reason about multi-step behaviors or to generate actionable detections. 
To address this gap, we propose to extract the layered Attack-driven Threat Intelligence (ATIs), a comprehensive representation that captures the full spectrum of adversarial behavior. We propose \name, which can accurately identify the ATIs including complete tactics, techniques, multi-step procedures, and their procedure variants, and integrate the threat intelligence to software security application scenarios: the detection rules (i.e., Sigma) and attack command can be generated automatically to a more accuracy level. 
Experimental results on 1,769 newly crawled reports and 16 manually calibrated reports show \name's effectiveness in identifying accurate techniques, outperforming state-of-the-art approaches of AttacKG by $1.34\times$ in F1 score.
Further studies upon 64,185 application logs via Honeypot show that our Sigma rule generator significantly outperforms several existing rules-set in detecting the real-world malicious events. 
Industry partners
confirm that our Sigma rule generator can significantly help save time and costs of the rule generation process. 
In addition, our generated commands achieve an execution rate of 99.3\%, compared to 50.3\% without the extracted intelligence.  
\end{abstract}


\maketitle

\section{Introduction}~\label{secintro} 
Advanced Persistent Threat (APT) attacks are serious attacks in the modern world, causing massive losses.  
APT attack behaviors often leave traces in various organization-internal log monitoring systems, such as system or web application logs.
To counter APT attacks, Security Operation Centers (SOCs) typically rely on Security Information and Event Management (SIEM) systems~\cite{splunk,DBLP:journals/ieeesp/BhattMZ14:SIEM, DBLP:journals/corr/abs-2311-10197:SIME-rule-evasion} to continuously collect and correlate these security events/logs, carrying out threat steps for real-time incident response. 
Beyond internal threat logs, external textual sources—such as blog posts (e.g., Cisco Talos~\cite{cisco-talos-cti}), government advisories (e.g., DARPA TC~\cite{Drapa-tc-e3}), open-source knowledge bases (e.g., \verb|MITRE ATT&CK|~\cite{mitre-attack}), security forums, and analyst reports—also provide Cyber Threat Intelligence (CTI) that highlights key attack elements such as techniques used.

Raw CTI reports are typically in natural language paragraphs, loosely structured and contain much redundant information, making them difficult to interpret and act upon. To facilitate rapid dissemination of threat understanding and effective cross-team collaboration, 
security analysts typically abstract the threat intelligence from the unstructured report, e.g., a hierarchical intelligence structure like Tactics, Techniques, and Procedures (TTPs).
Well-structured threat intelligence can enable and enhance several core security threat detections and understandings, including:

\begin{itemize}[fullwidth,itemindent=0em] 
    \item \textbf{Sigma Rule Configuration:} 
    The real-time demand for web security makes SIEM systems and their applied security rule an integral part of the intrusion detection life-cycle. 
    Sigma provides a standardized and vendor-agnostic format to define detection constraints of malicious events from chaotic logs within most SIEM systems (e.g., Splunk~\cite{splunk}, IBM QRadar). 
    An effective Sigma rule configuration relies on accurate threat intelligence, e.g., the \texttt{detection} and \texttt{tag} fields depend on the Procedures and Tactic-Technique pairs. 
    \item \textbf{Attack Behavior Validation:} 
    Generating reproducible attack scripts becomes crucial whenever analysts must recreate real threat behaviors to test defenses, validate rules, or ensure consistent benchmarking. Security experts and red teams often generate reproducible attack commands (e.g., typical spear-phishing PowerShell scripts), to simulate attack paths, understanding the attack dependencies and boundaries. The effective and correct commands is bounded to a series of  implementation steps (i.e., attack procedures). 
\end{itemize}

\subsection{Existing Methods and Limitations} 
We investigate existing researches for threat intelligence extraction  and present a consolidated summary in Table~\ref{tab:survery-threat-intelligence}. 
However, we summarize that prior works like TTPDrill~\cite{DBLP:conf/acsac/HusariAACN17:TTPDrill} or ThreatRaptor~\cite{DBLP:conf/icde/GaoSLXQXMKS21:enabling-hunt-ICDE21:THREATRAPTOR} have typically extracted a subset of individual threat intelligence—most commonly techniques—leaving out richer elements such as procedures, and procedure variational behaviors or explanations. This narrow focus limits both the depth and completeness of the resulting intelligence, making it difficult for software security systems to accurately reconstruct attack chains, assess adversary capabilities, or support high-fidelity detection engineering. Nowadays, real-world cyberattacks have recently evolved and often become complex, which can appear benign when viewed in isolation, while multi-step attack chains unfold across multiple related events.
Besides, STIX, the TTP extraction approaches (in Table~\ref{tab:survery-threat-intelligence}) and even recent efforts of CTINexus~\cite{cheng2025ctinexusautomaticcyberthreat:CTINExus} translating reports into knowledge graph representations, all assume that all threat intelligence are explicitly included in original report, neglecting implicit processes of attack variants. Such process variants, although not directly mentioned, are often key pathways used by advanced attackers to evade detection~\cite{DBLP:conf/ndss/OzmenSFC23:evasion-physical, DBLP:conf/uss/MukherjeeW0WCKK23:Evasion}. 

In addition, many existing approaches rely heavily on generic NLP pipelines, which often struggle with domain-specific terminology, subtle semantic cues, and cross-sentence dependencies. As a result, technique identification becomes noisy and error-prone, leading to incomplete or inconsistent coverage of real attacks. The inaccurate threat intelligence leads to the poor Rule/Command Adaptability. 
High-quality Sigma rules rely on accurate and comprehensive threat intelligence, particularly detailed knowledge of techniques and the procedures used to carry them out. Techniques provide the high-level “what” of an adversary’s behavior, allowing rule authors to target the correct tactic or behavioral pattern. However, procedures capture the concrete “how”—the specific commands, tool invocations, parameter choices, and execution sequences that adversaries actually use in practice. Without precise procedure-level intelligence, Sigma rules often become overly generic, miss critical variants, or fail to trigger on realistic attack traces. Conversely, when techniques and procedures are well understood, rule authors can design more discriminative conditions, reduce false positives, and ensure the rules cover real-world attack paths rather than only abstract tactics. 
As a result, accurate threat intelligence directly determines the effectiveness, robustness, and operational value of Sigma-based detection.

Together, these limitations reduce the practical usefulness of extracted CTI, hinder automation, and weaken the reliability of security analytics that depend on high-quality threat intelligence.

\begin{table*}[]
\centering
\setlength{\abovecaptionskip}{0pt}
\setlength{\belowcaptionskip}{0pt}
\caption{Approaches on threat intelligence extraction. NLP: natural language processing; IR: information retrieval.} 
\label{tab:survery-threat-intelligence}
\renewcommand\tabcolsep{1.2pt}
\begin{threeparttable}
\scalebox{0.93}{
\begin{tabular}{llcccccccccccll}
\toprule
                     &  & Tactic                 &  & Technique              &  & Procedure    & & Procedure Variant      & & Software Security Task Adaption & & Methods&  & Year \\ \cmidrule{1-1} \cmidrule{3-3} \cmidrule{5-5} \cmidrule{7-7} \cmidrule{9-9} \cmidrule{11-11} \cmidrule{13-13} \cmidrule{15-15}
iACE~\cite{DBLP:conf/ccs/LiaoYWLXB16:iACE}                 &  & $\times$   &  & $\times$   &  & $\times$ & & $\times$ & & $\times$ & & Regex  &  & 2016 \\ \cmidrule{1-1} \cmidrule{3-3} \cmidrule{5-5} \cmidrule{7-7} \cmidrule{9-9} \cmidrule{11-11} \cmidrule{13-13} \cmidrule{15-15}
TTPDrill~\cite{DBLP:conf/acsac/HusariAACN17:TTPDrill}            &  & $\checkmark$ &  & $\checkmark$ &  & $\times$  & & $\times$ & & $\times$ & & NLP and IR &  & 2017 \\ \cmidrule{1-1} \cmidrule{3-3} \cmidrule{5-5} \cmidrule{7-7} \cmidrule{9-9} \cmidrule{11-11} \cmidrule{13-13} \cmidrule{15-15}
rcATT~\cite{DBLP:journals/corr/abs-2004-14322:rcATT}                &  & $\checkmark$ &  & $\checkmark$ &  & $\times$   & & $\times$ & & $\times$ & & Supervised learning &  & 2020 \\ \cmidrule{1-1} \cmidrule{3-3} \cmidrule{5-5} \cmidrule{7-7} \cmidrule{9-9} \cmidrule{11-11} \cmidrule{13-13} \cmidrule{15-15}
EXTRATOR~\cite{DBLP:conf/eurosp/SatvatGV21:Extractor}             &  & $\checkmark$ &  & $\times$   &  & $\times$   &  
 & $\times$ & & $\times$ & & NLP & & 2021 \\ \cmidrule{1-1} \cmidrule{3-3} \cmidrule{5-5} \cmidrule{7-7} \cmidrule{9-9} \cmidrule{11-11} \cmidrule{13-13} \cmidrule{15-15}
ThreatRaptor~\cite{DBLP:conf/icde/GaoSLXQXMKS21:enabling-hunt-ICDE21:THREATRAPTOR}         &  & $\times$   &  & $\checkmark$ &  & $\times$   & 
& $\times$ & & $\times$ & & Unsupervised learning &  & 2021 \\ \cmidrule{1-1} \cmidrule{3-3} \cmidrule{5-5} \cmidrule{7-7} \cmidrule{9-9} \cmidrule{11-11} \cmidrule{13-13} \cmidrule{15-15}
AttacKG~\cite{DBLP:conf/esorics/LiZCL22:attackKG}             &  & $\times$   &  & $\checkmark$ &  & $\checkmark$   &  
& $\times$ & & $\times$ & & Knowledge graph & & 2022 \\ \cmidrule{1-1} \cmidrule{3-3} \cmidrule{5-5} \cmidrule{7-7} \cmidrule{9-9} \cmidrule{11-11} \cmidrule{13-13} \cmidrule{15-15}
CTINexus~\cite{cheng2025ctinexusautomaticcyberthreat:CTINExus}             &  & $\times$   &  & $\times$ &  & $\times$   &  
& $\times$ & & $\times$ & & LLMs & & 2025 \\ \cmidrule{1-1} \cmidrule{3-3} \cmidrule{5-5} \cmidrule{7-7} \cmidrule{9-9} \cmidrule{11-11} \cmidrule{13-13} \cmidrule{15-15}
\name &  & $\checkmark$ &  & $\checkmark$ &  & $\checkmark$ & & $\checkmark$ & & $\checkmark$ & & LLMs & & -- \\ \bottomrule
\end{tabular}
}
\caption*{\footnotesize $\ast$ Variant refers to a procedure that is new but plausible inference beyond the original report based on a grounded technical database}

\end{threeparttable}
\end{table*}

\subsection{Our Solution: \name}~\label{sec:threat-intelligence-extraction}    
Drawing from both a systematic review of prior works and our own experience of building and evaluating CTI pipelines, we arrive at building the whole threat intelligence analysis and integration.
First, we propose \textbf{attack-driven threat intelligence (ATI)} --a new structured format including the complete tactic, technique and procedure (TTP) chains, implicit procedure variants and explanations.   
The variants, which are although not explicitly mentioned in reports, but are grounded based on pluggable threat database, strengthening the detection coverage beyond observable behaviors~\cite{DBLP:journals/corr/abs-2311-10197:SIME-rule-evasion, DBLP:conf/uss/MukherjeeW0WCKK23:Evasion}. 
Second, with the ability of large language models (LLMs) to understand unstructured language and perform complex reasoning ~\cite{DBLP:journals/corr/abs-2305-18438:reinforcement-learning-genAI, DBLP:conf/aaai/LongZMMZZ024:GenAI, DBLP:journals/corr/abs-2409-02074:shell-command-explainer}, we deploy LLM-powered pipeline to improve the precision and depth of intelligence extraction, and the adaptability of rules and commands.
We integrate these components into \name — a system designed to extract ATIs and improve adaptability for software security tasks. \name addresses the limitations by: i) To improve the technique extraction accuracy, we combine a semantic \textit{chunker} algorithm, a combined component of \textit{Knowledge Enhancer}, and a \textit{Validator} to solve the problem of missed techniques and those incorrectly identified techniques. ii) ATI inference is guided by a mapping mechanism that integrates an external database with a graph, enabling \name to infer plausible behaviors beyond those explicitly described based on the connecting multiple pieces of threat information, linking dispersed evidence such as log entries, tool usage and targeted assets, keeping inferences grounded in external knowledge. 
iii) \name can leverage the proposed attack-driven intelligence to significantly enhance downstream software security tasks such as Sigma rule generation and command construction. It embeds a set of mapping functions that align the extracted ATIs—such as techniques, procedures, parameters, and contextual cues—with the corresponding fields required in Sigma rules and executable attack commands. This grounding allows \name to produce detection logic and commands that are not only syntactically correct but also semantically tied to real adversarial behaviors. The initial Sigma rules and attack commands then autonomously interact with syntax-checking modules (e.g., Sigma Lib, Python AST) and external live systems (e.g., Splunk~\cite{splunk}) to validate grammar, refine rule conditions, and adapt configurations to different environments. By iteratively enforcing correctness and system compatibility, \name transforms raw attack intelligence into operational artifacts that are tuned for accuracy, robustness, and real-world deployability.

\begin{figure*}
\centering
\scalebox{0.95}{
\includegraphics[width=0.94\linewidth]{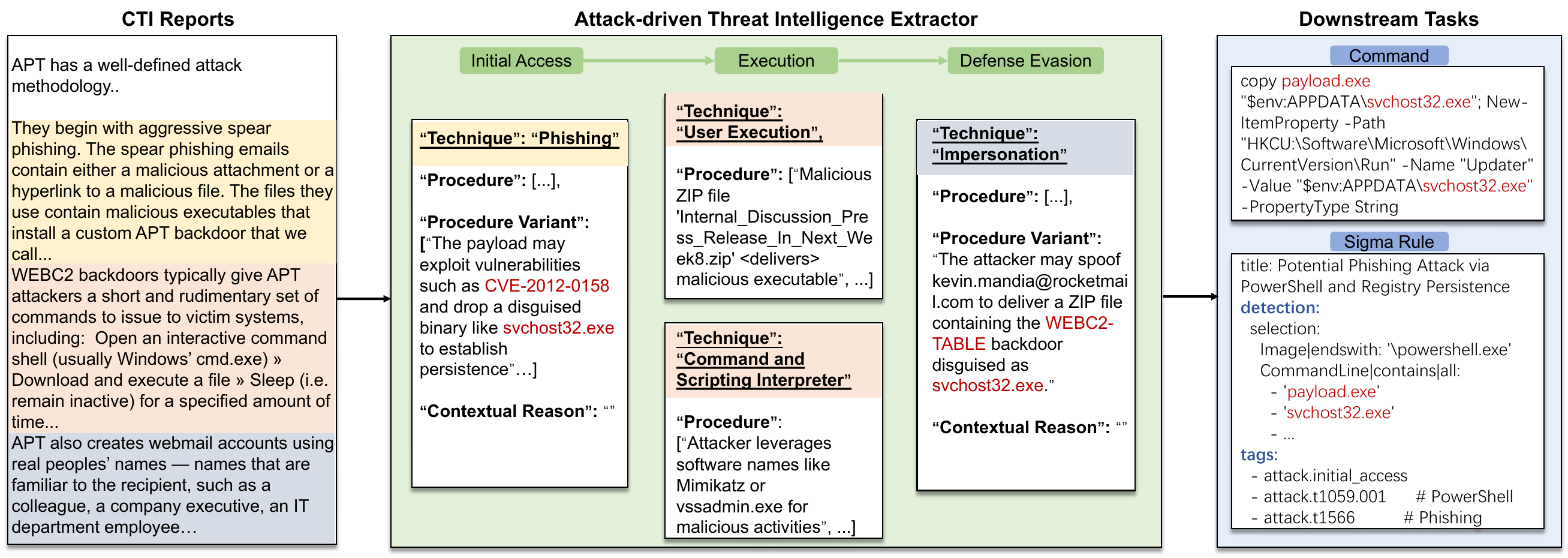}}
\caption{A demonstrating example of \name. The prior works show individual intelligence while \name exhibits the whole attack-driven intelligence including three attack stages (e.g., Initial Access), with their techniques (e.g., Phishing), procedures and variants, and contextual explanations. The Sigma rule and command are derived from the ``phishing'' technique.}
\label{fig:motivation}
\end{figure*}



\noindent\textbf{A Demonstrating Example.} 
We show a demonstrating example in Figure~\ref{fig:motivation} when interpreting the snippet of a CTI report~\cite{CTI-used} into our-proposed ATIs. 
For example, while the original CTI report may describe a phishing email leading to malware installation, our system can infer that the malware exploits known vulnerabilities such as \texttt{CVE-2012-0158} and drops a disguised binary like \texttt{svchost32.exe} to establish persistence.

The extracted ATI provides richer, procedure-level semantics that align naturally with what downstream software security tasks require, enabling more accurate and context-aware rule and command generation. Sigma rules~\cite{Sigma} are fundamentally pattern specifications of suspicious or malicious behavior; they rely on precise field–value mappings—such as matching execution entities, process command lines, or behavioral sequences—to function effectively. However, these field values are not arbitrary strings: they must faithfully correspond to concrete attacker actions, tools, and parameters described in real attack procedures.
For example, the detection field contains the core functional conditions, which are mapped from the attack execution procedures: the \texttt{Image} matches the execution entity, and the \texttt{CommandLine} matches the procedures (from the phishing techniques). 
Other fields might include the title as a brief explanation of what the rule detects, the tags as techniques classifications. 
In this way, \name leverages ATI to bridge a long-standing semantic gap between high-level threat intelligence and the low-level configuration values required by Sigma rules. This mapping ensures that generated rules not only follow syntactic correctness but also faithfully capture the operational patterns of the attack, leading to more robust, precise, and environment-aware detection logic for downstream software security systems.

\noindent\textbf{Results.} 
We deploy \name upon GPT-4o,
GPT-4o-mini, 
DeepSeek-V3, and LLaMa3.
Experimental results based on 1,769 recently crawled reports~\cite{cisco-CTI-report} and 16 manually calibrated reports~\cite{AttackKG-ground-truth} show that all deployed \name can accurately identify techniques, outperforming the state-of-the-art approach of AttacKG~\cite{DBLP:conf/esorics/LiZCL22:attackKG} by up to $1.34\times$ in F1 scores.   
Based on 64,185 application logs collected by a Honeypot in the wild and a Splunk SIEM for rules' executing, we show that the generated Sigma rules by \name can catch significantly more malicious events than that by standalone LLMs and the existing rules-set from Sigma~\cite{Sigma} and Splunk~\cite{Splunk-rules}. 
We evaluate the generated commands to verify their executability: the commands by \name achieves up to 99.3\% execution rate, which is significantly higher than the 50.3\% of commands generated without procedures.

We summarize our contributions as follows. 

\begin{itemize} 
    \item \textbf{Attack-driven Threat Intelligence.} We propose a novel representation of ATI, which consists of whole TTPs and the procedure variants with explanations.
    We design \name to accurately extract ATIs, which are used to guide the generation of Sigma rules and commands. 
    \item \textbf{Extension Evaluation.} We evaluate \name upon large-scale CTI reports, demonstrating its higher ATI extraction accuracy and down-stream tasks' adaptability. Tested upon real-world logs, the Sigma rules generated from \name can significantly outperform that generated from general LLMs, and the static rules-set in detecting live malicious events. 
    \item \textbf{Real-world deployment.} We deploy our generated Sigma rules upon real-world web application logs collected via a honeypot, showcasing their better effectiveness that existing rule-sets in detecting real-world malicious events. 
\end{itemize}

\noindent We commit to release all our used datasets and codes for reproducibility of data and results. 

\section{~\label{sec:framework}Design of \name}

\begin{figure}[t]
\centering
\scalebox{0.5}{
\includegraphics[width=\linewidth]{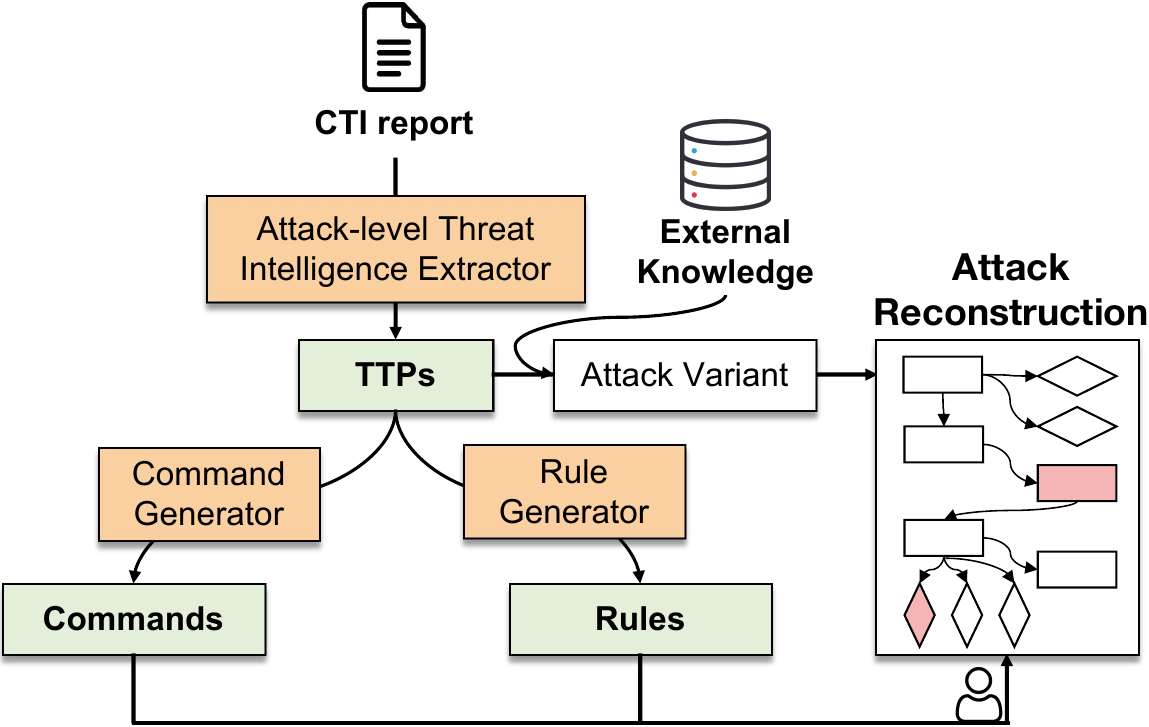}}
\caption{General overview of \name, which includes ATI extractor, command generator, and rule generator}
\label{fig:general-overview}
\end{figure}

\noindent\textbf{Overview of \name.} Given an unstructured CTI report as input, \name (shown in Figure~\ref{fig:general-overview}) first extracts structured attack-driven threat intelligence (ATI), including TTPs and attack variants, which are then fed into \name to generate Sigma rules and reproducible commands.


\subsection{Attack-driven Threat Intelligence Extractor}

Given an unstructured report $\mathcal{R}$, and a technique database $\mathbb{T} = \{(t_i, d_i)\}_{i=1}^N$ consisting of $N$ techniques $t_i$ and their corresponding standard descriptions $d_i$.
The first goal is to extract a subset of relevant techniques $\mathbb{T}_\mathcal{R}$ mentioned in $\mathcal{R}$.

\begin{equation}\nonumber
\mathbb{T}_\mathcal{R} \subseteq \{t_i\}_{i=1}^N
\end{equation}

Given the extracted techniques $\mathbb{T}_\mathcal{R}$, the original report $\mathcal{R}$, and an external procedure database $\mathcal{P}$, our goal is to generate a set of procedures and their variants using a generation function. 

\begin{equation}\nonumber
\hat{P} = \{p_j\}_{j=1}^M 
\end{equation}

\noindent We follow the top-to-down strategy with a stepwise execution, sequentially outputting each intelligence by using the output of $o_i$ (i.e., tactics-techniques) as contextual input for $o_{i+1}$ (i.e., procedures), to ensure an progressive intelligence extraction.   
We use the technique database from \verb|MITRE ATT&CK|~\footnote{\url{https://attack.mitre.org/} \label{footnote:mitre}}, as it serves as a standardized reference for the security community, providing a globally accessible taxonomy of real-world adversary tactics and techniques.

\subsubsection{<Tactic, Technique> Identification} 
\begin{figure*}
\centering
 \includegraphics[width=0.93\linewidth]{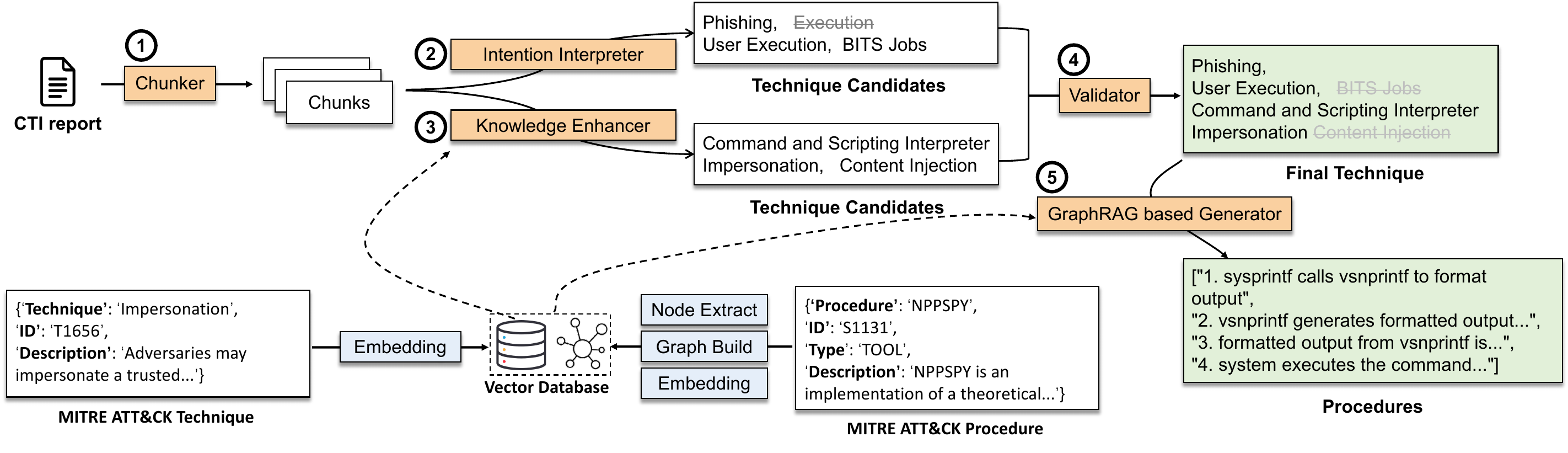}\caption{Overview of attack-driven threat intelligence (ATI) extractor.} 
\label{fig:overview}
\end{figure*}

As shown in Figure~\ref{fig:overview}, to enable an accurate identification, we first propose a chunker component~\textcircled{1} to decompose $\mathcal{R}$ into multiple smaller chunks, which are expected to contain attack behaviors. 
The chunks are passed into an \textcircled{2} intention interpreter to extract the technique candidates via the in-context learning abilities of LLMs, followed by a \textcircled{3} knowledge enhancer to strength more techniques via a external domain-specific database, concluded by  \textcircled{4} validator to judge the technique candidates generated by previous steps as the final recognized techniques.  




\noindent\textbf{CTI Chunker.} 
We first propose a CTI chunker that segments a long report into smaller and meaningful chunks, with the goal of ensuring that a chunk retains critical intelligence, eliminating irrelevant information. 
This is because that LLMs face inherent performance limitations with long inputs due to inefficiencies in the attention mechanism and information loss across long sequences~\cite{Press2022TrainShort, Liu2023LostInTheMiddle, DBLP:journals/corr/abs-2502-12962:LLMwindows, DBLP:conf/ijcai/WangSORRE24:LLMwindows}.

As outlined in Algorithm~\ref{algo:chunking}, our algorithm first identifies that those blocks (e.g., $N$ sentences) containing IoC (Indicators of Compromises) entities, such as malicious IP addresses or suspicious file hashes. We then merge the surrounding (i.e., preceding and following) sentences with the blocks as the final chunks, with the granularity controlled by the parameters of $block_{size} (N)$ and $MergedSentence_{size}$. The core principle is that IoCs represent fundamental indicators of malicious activity and encapsulate critical threat intelligence; blocks lacking IoCs are likely irrelevant to actual threats.
To extract IoCs, we design a few-shot prompt, with specified tasks stating that the task is to extract IoCs such as IP addresses, domains, URLs, file hashes, and email addresses. 
We don't consider traditional methods such as regular expressions, pattern matching, dictionary-based lookups for extracting IoCs, as they always lack semantic understanding, leading to false positives, e.g., extracting file hashes simply based on their format, without considering surrounding context.

\begin{algorithm}[t]
\footnotesize
\caption{\label{algo:chunking} <Tactic, Technique> Identification} 
\begin{flushleft}
\textbf{Input:} CTI report $\mathcal{R}$; vector database $\mathcal{D}$ \\
\textbf{Output:} Identified <tactic, technique> pairs $\mathbb{T}_R$ 
\end{flushleft}
    \begin{algorithmic}[1]
        \STATE \textbf{// Step 1: CTI Chunking}
        \STATE $\mathcal{C} \leftarrow \{\}$  // Chunked report set
        \STATE Split $\mathcal{R}$ into $m$ blocks ${b_1,\dots, b_m}$, each with $N$ sentences
        
        \FOR{each $b_i \in \{b_1, ..., b_m\}$}
            \IF{LLM\_ContainsIoC($b_i$)}
                \STATE $r_i \leftarrow$ MergeSentencesAround($b_i$, $MergedSentence_{size}$) \hfill // Expand block to include surrounding context
                \STATE $\mathcal{C} \leftarrow \mathcal{C} \cup \{r_i\}$ \hfill // Add chunk $r_i$ to the valid chunk set
            \ENDIF
        \ENDFOR
        \STATE \textbf{// Step 2: Intention Interpreter}
        \FOR{each $r_i \in \mathcal{C}$}
            \STATE $q_i \leftarrow$ CoT\_Prompt($r_i$) \hfill // Generate reasoning output using CoT
            \STATE $\hat{\mathbb{T}}_{r_i}^{\text{CoT}} \leftarrow$ Parse\_CoT($q_i$) \hfill // Parse predicted tactic-technique pairs
        \ENDFOR
        \STATE \textbf{// Step 3: Knowledge Enhancer based on Vector Database Retrieval}
        \FOR{each chunk $r_i \in \mathcal{C}$}
            \STATE $E_{r_i} \leftarrow \mathcal{V}(r_i)$ \hfill // Compute embedding vector of chunk $r_i$
            \STATE $\mathcal{S}_i \leftarrow \{\mathcal{S}(E_{r_i}, \mathcal{V}(d_j)) \mid d_j \in \mathcal{D} \}$ \hfill // Compute similarity scores
            \STATE $\hat{\mathbb{T}}_{r_i}^{\text{Vec}} \leftarrow$ TopK($\mathcal{S}_i$, $k$) \hfill // Select top-k most similar techniques
        \ENDFOR
        \STATE \textbf{// Step 4: Validation and Consolidation}
        \STATE $\mathbb{T}_R \leftarrow \emptyset$ \hfill // Initialize final result set
        \FOR{each $t_i \in \bigcup_i \left(\hat{\mathbb{T}}_{r_i}^{\text{CoT}} \cup \hat{\mathbb{T}}_{r_i}^{\text{Vec}}\right)$}
            \STATE $d_i \leftarrow$ GetDescription($t_i$) \hfill // Retrieve standard MITRE description
            \STATE $verdict_i \leftarrow F(d_i, \mathcal{R})$ \hfill // Use independent LLM-as-a-judge to assess semantic alignment
            \IF{$verdict_i = true$}
                \STATE $\mathbb{T}_R \leftarrow \mathbb{T}_R \cup \{t_i\}$ \hfill // Retain the technique
            \ENDIF
        \ENDFOR
        \RETURN $\mathbb{T}_R$
    \end{algorithmic}
\end{algorithm} 

\noindent\textbf{Intention Interpreter.} 
With the segmented chunk-list $\mathcal{C} = \{r_1, r_2,...,r_i\}$, we employ a two-stage intention interpreting framework by first performing \textit{Tactic Planning} to identify high-level attacker objectives, followed by \textit{Technique Generation} to derive specific techniques. 
The Chain-of-Thought (CoT) reasoning process can be formally represented as 

\begin{equation} \nonumber 
\mathcal{T}_{r_i} \leftarrow \text{Tactic\_Plan\_Prompt}(c_i) 
\end{equation}

\noindent where $\text{Tactic\_Plan\_Prompt}$ is a guided CoT prompt designed to elicit reasoning about attack intent, constrained to the 14 official tactics in \verb|MITRE ATT&CK|. The LLM is instructed to infer a plausible tactic plan based on the semantics of the input chunk. 
For each identified tactic $\tau \in \mathcal{T}_{r_i}$, we further generate candidate techniques that are conditioned on both the original chunk and the given tactic:

\begin{equation} \nonumber
\hat{\mathbb{T}}_{r_i}^{\tau} \leftarrow \text{Guided\_Generation}(c_i, \tau) \end{equation}

\noindent where $\text{Guided\_Generation}$ incorporates a hybrid processing including LLM generation and rule-based filtering. We leverage a curated set of tactic-specific rules that link tactics to indicative keywords or behavioral patterns. If a sentence in the chunk matches the associated rules, we trigger a tactic-aware prompt to generate relevant techniques. 

To mitigate hallucination and ensure factual alignment with \verb|MITRE ATT&CK|, we perform a post-processing step to validate the generated techniques. Each technique is checked against the official technique list. If a generated technique is not found, we compute its embedding and retrieve the top-3 most similar techniques within the same tactic. A replacement is made if a semantically aligned technique is identified; otherwise, the hallucinated output is discarded.



\noindent\textbf{Knowledge Enhancer.}  
To enhance the extraction of missed techniques, we incorporate a retrieval-augmented generation (RAG) strategy~\cite{DBLP:conf/nips/LewisPPPKGKLYR020:RAG-first} that dynamically compare the standard technique descriptions with the input chunks.
The RAG knowledge base is bootstrapped from MITRE's official documentation~\cite{mitre-attack}, which provides detailed illustrations for the technique's meaning. From this corpus, we build the the vector database storing key-value pairs of <tactic-technique: description>, where descriptions are projected into an embedding space for comparison of similarity with input chunks. 
Given that lengthy standard descriptions (e.g., spanning three or four paragraphs) may introduce information loss, we calculate the average word count of the first paragraph and use this value (i.e., 128 tokens) to segment the standard descriptions into several $d_j$. Finally, the chunks are embedded and the dot product is calculated with all the vectors in vector database.

In this way, more techniques can be identified by 

\begin{equation}\nonumber
\hat{\mathbb{T}}_{r_i}^{\text{Vec}} = \mathop{\arg\max}\limits_{i \in \mathbb{T}} \operatorname{avg} \left( Top_k[\mathcal{S}(\mathcal{V}(r_i), \mathcal{V}(d_j)]\right)
\end{equation}

\noindent where $\mathcal{V}(\cdot)$ denotes the 
vector embedding model (\texttt{text-embedding-ada-002}~\cite{openAI-vector-model} in our task), $S(\cdot,\cdot)$ denotes the function to compute the similarity score, $\mathbb{T}$ is the set of all techniques, i.e., 235 techniques in \verb|MITRE ATT&CK|. The $Top_k(\cdot)$ is the function that picks the $k$ similar techniques, which are returned as the identified technique candidates. 

\begin{table*}[h!]
\setlength{\abovecaptionskip}{0pt}
\setlength{\belowcaptionskip}{0pt}
\caption{Prompt template for technique classifications. 
Our prompt is structured into four key components: Background, Task, Guidelines, and optional examples as a progress approach for analyzing, refining, and integrating information. \label{tab:prompt-template}}  
\centering
\footnotesize
\begin{tabularx}{\linewidth}{X} 
    \toprule[\thickline]
    \coloredbox{lightpurple}{Background:}
    You are a helpful assistant with the security domain knowledge. \\ 
    \midrule[\thinline]    
    \textbf{IoC Recognition for <CTI Chunker}>: \\
    \coloredbox{lightyellow}{Task:} 
    Your task is to extract the indicators of compromise (IoCs) from the following CTI report. \\ 
        \coloredbox{SoftGreen}{Examples:}
    IP Address, Domains, URLs, Email Addresses; 
    File Hash, URL, Process Names, Registry Keys. \\
    \textbf{Final prompt:} 
    \coloredbox{lightpurple}{Background} + \coloredbox{lightyellow}{Task} + \coloredbox{mygray}{Guidelines} + \coloredbox{SoftGreen}{Examples} + \underline{\$Chunks | Report}
    \\
    \midrule[\thinline]
    \textbf{<Intention Interpreter}>: \\
    \coloredbox{lightyellow}{Task:}
    Identify the high-level ATT\&CK tactics that the threat actor is likely pursuing in the order of execution. Use only tactics from the official \texttt{MITRE ATT\&CK} list.
    Below are the tactics' names and descriptions:
    \underline{\{TACTIC\_DESCRIPTION\}}\\
    Based on the tactic, list the \texttt{MITRE ATT\&CK} techniques this behavior likely represents. Below is the complete mapping of each tactic to its associated techniques: 
    \underline{\{TACTIC\_TECHNIQUES\_MAPPING\}}\\
    \textbf{Final prompt:} 
 \coloredbox{lightpurple}{Background} + \coloredbox{lightyellow}{Task} + \coloredbox{mygray}{Guidelines} + \underline{\$Chunks | Report} 
    \\

    \midrule[\thinline]
    \textbf{<Knowledge Enhancer}>: \\
     \coloredbox{lightyellow}{Task:}
    According to the report and the vector database, find the techniques most relevant to the description\\
    \textbf{Final prompt:} 
     \coloredbox{lightpurple}{Background} + \coloredbox{lightyellow}{Task} + 
     \underline{\$Chunks | Report}
     \\ 
    \midrule[\thinline]
   \textbf{<Validator}>: \\
    \coloredbox{lightyellow}{Task:} 
    The user will provide another possible technique and description.
 You need to verify whether the technique exists in the report.\\
    \textbf{Final prompt:} 
     \coloredbox{lightpurple}{Background} + \coloredbox{lightyellow}{Task} + \coloredbox{mygray}{Guidelines} +
     \underline{\$Technique Candidates: Standard Descriptions}
     + \underline{\$Report}
     \\
    \bottomrule[\thickline] 
\end{tabularx}
\end{table*}

\noindent\textbf{Validator.} 
The technique candidates generated in previous two steps might be irrelevant to the report, leading to false positives. 
To mitigate this, we introduce a validator to ensure that most responsible and correct techniques are finally settled down. 
Specifically, validator employs a custom LLM, isolated from prior context, to judge the technique candidates, removing the overly general techniques.   
The validator follows the grounded reasoning by re-thinking and comparing two descriptions between the entire report $\mathcal{R}$ and the descriptions of every technique candidates $d_i$, as formalized below. 

\begin{equation}\nonumber
T_{\textit{final}} = \{ t_i \in T_{\textit{candidates}} | F(d_i, R) \implies true \} 
\end{equation}

\noindent where $F(\cdot,\cdot)$ is a function to distinguish whether the semantics and intention align between two texts. 
The function $F(\cdot,\cdot)$ is achieved by a validator prompt with detailed guidelines.
Note that the LLM used in \emph{LLM-as-a-judge} operates independently, with no shared memory, and is assigned different system messages to ensure they remain focused on their specific subtasks.



\subsubsection{Procedure Generation} 
To enable the capabilities of modeling relationships between entities for procedures and their variants, we employ GraphRAG~\cite{DBLP:journals/corr/abs-2404-16130:graphRAG}, which specializes in the joining operations via the built-in graph structure. Instead of the RAG, we specifically design a multi-hop GraphRAG-based reasoning modular to estimate the relationships among entities.
To achieve this,  \name first performs a report narrowness to find a meaningful part (i.e., procedure paragraph) given the technique guidance. The procedure paragraph is linked through the snippet of external vector database and the original report via reasoning; second, \name decouples the procedure paragraph into procedures. 
Formally, given a CTI report $\mathcal{R}$, the procedure generation is defined as:

\begin{equation} \nonumber
\hat{P} = \text{Gen}\left(Integrate(\mathcal{R}_{\text{ent}}, \mathcal{R}_{\text{rel}}, \mathcal{R}_{\text{comm}}, \mathcal{R}_{\text{sum}}) \right)
\end{equation}

\noindent where the fused context combines four types of retrieved information: entity-level mentions ($\mathcal{R}_{\text{ent}}$), relationship paths ($\mathcal{R}_{\text{rel}}$), community clusters ($\mathcal{R}_{\text{comm}}$), and high-level summaries ($\mathcal{R}_{\text{sum}}$). These are jointly used by the generation module $\text{Gen}(\cdot)$ to produce the final procedures $\hat{P}$.


\name dismantles the procedure paragraph (shown below) into several procedures by reasoning the relevant entities and relationships, where the procedure paragraph is referenced from the external reports with their array numberings (as exampled below).
When the model encounters a partially described procedure in the CTI report, it leverages the graph-based retrieval to infer related steps, tools, or IoCs by identifying similar entities and neighborhood structures. \name matches extracted IoCs and entities in the report with nodes in the knowledge graph and retrieve their adjacent nodes (e.g., common follow-up actions or co-occurred artifacts).  
This guided retrieval allows the model to generate faithful variants that may not be explicitly mentioned in the text but are supported by the knowledge database, rather than hallucination from the LLMs.   
\name uses a classification function $f_{\text{variant}}(\hat{P}, \mathcal{R})$ to identify whether a generated procedure $\hat{P}$ constitutes a novel \textit{attack variant}, 
providing multiple levels of similarity matching to assess whether $\hat{p}$ can be traced back to $\mathcal{R}$. 
\name segments $\mathcal{R}$ into a sentence set

\begin{equation} \nonumber
\mathcal{S} = \{s_1, s_2, ..., s_n\}
\end{equation}

\noindent then, encode each sentence into an embedding set 

\begin{equation} \nonumber
\mathcal{E}_\mathcal{R} = \{e_1, e_2, ..., e_n\}
\end{equation}

\name employs normalized string matching, semantic similarity based on sentence embeddings, and keyword overlap. A procedure is classified as an attack variant if it cannot be sufficiently matched to any sentence in the report. We consider a procedure as matched if it either has high lexical overlap, exceeds a semantic similarity threshold (e.g., 0.84), or appears nearly verbatim in the text. 

\begin{figure}[t]
\footnotesize
\setlength{\abovecaptionskip}{0pt}
\setlength{\belowcaptionskip}{0pt}
\begin{tcolorbox}[colback=white!95!black,colframe=black!90!white,title= An example of the procedure generation process, fonttitle=\bfseries, sharp corners=south]
\textbf{Technique:} Command and Scripting Interpreter\\ 
\textbf{Procedure Paragraph}: The primary technique utilized in the attack is the **Command and Scripting Interpreter**, specifically through the use of **Windows' cmd.exe**. 
The use of cmd.exe is a common tactic in cyber attacks, enabling attackers to leverage built-in system functionalities for malicious purposes [Data: Reports (590, 728, 501, 463, 520)].  \\
\textbf{Procedure}: cmd.exe opens interactive command shell
\end{tcolorbox}
\label{fig:procedure}
\end{figure}

\noindent\textbf{Parameters.} 
The vector database consists of 677 procedures in \verb|MITRE ATT&CK|, in the format of $\langle \textit{technique: procedures} \rangle$. 
We input the whole report into GraphRAG, because it has the internal chunking mechanisms. 
The generated knowledge graph comprises 6,002 entities and 10,900 relationships within 36 categories. A category refers to a set of entities and paragraphs
with semantically consistent and contextual relevance. 

\subsection{Generator for Sigma Rules and Commands}
\begin{figure}[t]
\centering
\setlength{\belowcaptionskip}{-4pt}
\scalebox{1}{
\includegraphics[width=0.8\linewidth]{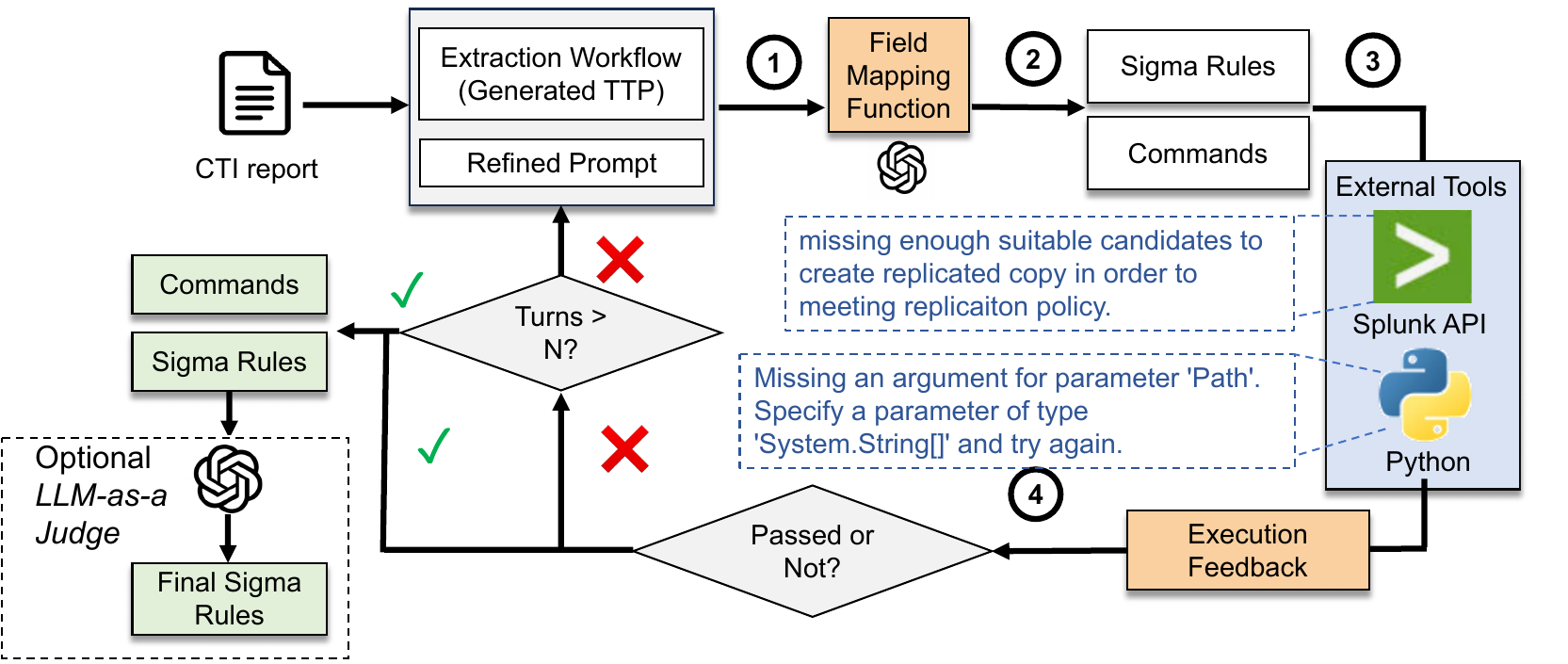}}
\caption{Overview of Sigma Rule/Command Generator.}
\label{fig:overview:rule-gen}
\end{figure}

\name integrates the ATI as the initial CoT reasoning prompts as the primary semantic driver, and then combine a system integration module linking external tools (e.g., Splunk API or Python AST), enabling iterative feedback and refinement (shown in Figure~\ref{fig:overview:rule-gen}) to generate Sigma rules and commands.  

\subsubsection{Sigma Rule Generator} 
\name employs the functional mapping that the specific parameter value of every fields (keywords) in a Sigma rule is purposely guided and grounded in corresponding ATIs. 
The core detection field is fully grounded in the \textit{procedure} or \textit{procedure variant} description, with field-value pairs reflecting key entities such as command lines, image names, registry paths, and network indicators.
For example, if the procedure specifies \texttt{powershell.exe -enc ...} to establish persistence, then the rule’s detection should reflect conditions on:
\texttt{Image=powershell.exe}, \texttt{CommandLine contains '-enc'}. We employ a semantic prompt with the template in Table~\ref{tab:prompt-template-rule-detection} to achieve the mapping functions above, where the rules can be tailored in Guidelines when the down-stream tasks unavoidably should include domain-specific or application-specific symbols.

To maintain the syntax and semantic consistencies, \name dynamically invokes modules via Splunk API to address identified issues when converting our generated Sigma rules into Splunk, chosen for its popularity~\cite{splunk,Splunk-rules}.   
Formally, let
  \( R = \{r_1, r_2, \dots, r_n\} \) represent a set of Sigma rule candidates. 
 \( f_{\text{yaml}}(r) \) be the function that checks if the rule \( r \) is in valid YAML format. \( f_{\text{sigma}}(r) \) be the function that checks if rule \( r \) follows Sigma syntax.
  \( g(r) \) be the function that converts rule \( r \) to a Splunk rule. \( f_{\text{splunk}}(g(r))\) be the function that checks the syntax of the corresponding Splunk rule.
  \( M() \) represents message LLMs used to regenerate a rule when validation fails. \name applies a series of the following reflection functions to iteratively refine the generated Sigma rules, ensuring grammatical correctness and enhancing semantic accuracy, with the goal of reducing false positives and improving true positive rates.

\begin{align*}
\text{YAML Format: }&  f_{\text{yaml}}(r) = \text{False}  \implies r' = M() \\  
\text{Syntax Check: }  & f_{\text{yaml}}(r) = \text{True} \land f_{\text{sigma}}(r) = \text{False} \\ &  \implies r' = M() \\  
\text{Splunk Rule Syntax: } &   f_{\text{yaml}}(r) = \text{True} \land f_{\text{sigma}}(r) \\ & = \text{True} \land f_{\text{splunk}}(g(r))  = \text{False} &\\ & \implies r' = M() \\  
\text{Final Condition: }&  f_{\text{yaml}}(r') \land f_{\text{sigma}}(r') \land f_{\text{splunk}}(g(r'))
 \end{align*}

\noindent The feedbacks contain syntax errors, missing field references, invalid logical conditions, or incomplete value mappings encountered during the translation of Sigma rules into Splunk-compatible queries (e.g., ``missing enough suitable candidates to create replicated copy in order to meeting replication policy'').

To ensure the precise rules, a validator is additionally employed to to filter out overly generic rules leading much excessive false positives. 
We empirically observed that rules like \texttt{Image IN (*CalculatorApp.exe, *hollow\_process\_name.exe)}, 
which detect whether the current process matches \textit{CalculatorApp.exe} or \textit{hollow\_process\_name.exe}, can be syntactically correct but overly generic, because such rules rely solely on the image field of the process name without considering other critical characteristics such as the file path, parent process, or command-line arguments. 
\name guides such semantic-misalignment by evaluating the relevance of the Sigma rules with respect to the entire CTI reports $R$.

\noindent\textbf{Environment.} We use the syntax validation using Splunklib's dry-run mode~\cite{githubSplunksdkpythonsplunklibMaster}, allowing to check for grammar errors without executing the query. 
We use the Sigma-CLI libraries~\cite{Sigma} to convert Sigma rule to Splunk rules. 
We set an attempt limit of reflection to 30 as the maximum number, discarding rules that exceed this threshold, avoiding an endless loop.

\begin{table}[t]
\setlength{\abovecaptionskip}{0pt}
\setlength{\belowcaptionskip}{0pt}
\caption{Prompts for Sigma rule generator.  
}    
\label{tab:prompt-template-rule-detection}
\centering
\footnotesize
\begin{tabularx}{\linewidth}{X}
    \toprule[\thickline]  
    <\textbf{Candidate Sigma Rule Generation}>: \\
     \coloredbox{lightyellow}{Task:} 
    For each attack description, write several Sigma rules to detect the relevant attacks.  
    The input is: 
    \$\underline{Report}.
    ATIs: \{\underline{\$ATIs}\}
    \\
        \coloredbox{mygray}{Guidelines}: Use the following mappings to transform each procedure and its context into a valid Sigma rule:\\
1. Procedure command / IoC → detection: Populate the detection field with key indicators such as file names, process paths, or network artifacts.\\
2. Tactic / Technique → tags: Map to tags to describe the detection scope (e.g., attack.execution, attack.t1059).\\
3. Procedure severity / tactic impact → level
Assign rule severity (e.g., low, high) based on the threat level implied by the tactic.\\
4. CTI source → references: Include links to the original CTI report or known vulnerabilities (e.g., CVEs) for traceability. \\
    \midrule[\thinline]  
    <\textbf{LLM-as-a-judge}>: \\
    \coloredbox{lightyellow}{Task:} 
    Your task is to filter out irrelevant rules from the generated rules.
    You will be given content from a  CTI report and a rule. You need to determine whether the rule is relevant to the CTI report.
    The cyber report is:
    \$\underline{Report}.
    The rule you need to judge: \{\$\underline{Sigma\_rule}\}
    \\
    \coloredbox{mygray}{Guidelines}: 
    1. Focus on alignment between the rule’s detection field and the report’s described actions. Ignore generic or unrelated rules. \\
    2. You should be very tough on the rules. If the rule is not relevant to the CTI report, you should filter it out. \\
    \bottomrule[\thickline]
\end{tabularx}
\end{table}

\subsubsection{Command Generator} The procedures and their variants are inherent nature of explicitly detailing the actions to be executed, aligning with the commands. 
\name maps the \texttt{type} field of a procedure to the corresponding execution environment, and extract the \texttt{command} field as the core of the generated executable instruction. When procedure contains placeholders (e.g., \texttt{{payload}} or \texttt{{target}}), \name substitutes them using predefined templates or contextual parameters derived from the original CTI report. 
For refinement\name goes through an external reflection module, interacting with the Python function feedback.  
The pipeline can be effectively simplified as:

\begin{equation}\nonumber
\mathcal{R} \xrightarrow{f_{\text{proc}}} P \xrightarrow{f_{\text{cmd}}} C\_{\text{candidates}} \xrightarrow{f_{\text{reflection}}} C
\end{equation}

\noindent where
\( P \) is intermediate procedure, and \( C \) is final command. We do not apply \emph{LLM-as-a-judge} to command generation, as commands are less prone to semantic ambiguity. 

\noindent\textbf{Environment.} We choose the platform compatible with the Windows environment, including PowerShell, CMD, Bash, Python, and other supported interpreters. 
We show CoT prompting in the command generation step 
in Table~\ref{tab:prompt-command-generator-procedures}. 

\begin{table}[t]
\setlength{\abovecaptionskip}{0pt}
\setlength{\belowcaptionskip}{0pt}
\caption{Prompts for command generator.}  
\label{tab:prompt-command-generator-procedures}
\centering
\footnotesize
\begin{tabularx}{\linewidth}{X}
    \toprule
    \coloredbox{lightyellow}{Task:} 
    Generate a valid, executable Windows command based on a given procedure description. The input is: 
    \$\underline{Report}.
    ATIs: \{\underline{\$ATIs}\}
    \\  
    \coloredbox{mygray}{Guidelines}: Use the following mappings to transform the procedure into a valid, executable Windows command:\\
1. Procedure type → Execution Environment:
Selects the appropriate shell (e.g., PowerShell, cmd).\\
2. Tactic / Technique → Behavior Check:
Ensure the command behavior aligns with the described attack goal.\\
3. CTI report context → Additional Parameters: 
Use extra context (e.g., file name, domain) to complete or customize the command. 
\\

    \bottomrule
\end{tabularx}
\end{table}





\section{Implementation and Evaluation Setup}

\subsection{Attack-driven Intelligence Extractor}

\noindent\textbf{Implementation Details.} 
We implement a prototype of \name in Python and open-source it for the community and integrate them into the client-side tool~\footnote{\url{https://threatpilot-web.streamlit.app/}} using the Streamlit library. 
We build \name upon GPT-4o, GPT-4o-mini, DeepSeek-V3 (671B)~\cite{DeepSeek-download} and LLaMa-3 (405B)~\cite{Llmama3} to show the \name's adaptability to various LLM backbones.
To control generation behavior, we set all LLMs to a temperature of 0.2 for a balance between determinism and flexibility, top-p of 0.1 for controlled diversity, frequency and presence penalties at 0, and a maximum response length of 512 tokens to prevent excessively long outputs.

In algorithm~\ref{algo:chunking}'s \emph{chunker}, we set $block_{size} (N)$ as 3 sentences and $MergedSentence_{size}$ as 1 sentence, striking a balance between processing speed and meaningful segmentation.
For balancing the false positives and false negatives, we set the top $k$ retrieval results from the vector database as 3. 

\noindent\textbf{AttacKG Implementation.} 
We reproduced AttacKG with its released code and model checkpoints, 
leveraging pre-trained Named Entity Recognition models for entity extraction. 
Specifically, we used spaCy’s $\verb| en_core_web_sm-3.1.0|$ model, fine-tuned on sampled \verb|MITRE ATT&CK| examples, with an optimizer of rule-based entity recognizer. 
For attack technique identification, AttacKG utilized graph alignment with a fuzzy matching algorithm, where the node alignment threshold was set to 0.65, and the graph alignment threshold was set to 0.85. 

\noindent\textbf{CTI Report Collection and Labeling.} 
First, we use the open-sourced experimental reports from AttacKG~\footnote{\url{https://github.com/li-zhenyuan/Knowledge-enhanced-Attack-Graph/tree/main/Results}. 
The gray and green parts together serve as the ground truth. 
}: 16 reports, including 5 attack reports from DRAPA TC and 11 reports regarding widely-occurred attack campaigns.
Due to the large-scale updates on \verb|MITRE ATT&CK| (e.g., 81.5\% of techniques' descriptions are modified since 2022), many techniques now differ significantly from previous versions~\cite{DBLP:journals/corr/abs-2409-02074:shell-command-explainer} (e.g., The technique of \texttt{Command and Scripting Interpreter} added scripting languages of PowerShell, Bash, Python than before), causing the original labeled ground truth to become outdated and lose effectiveness over time. 
To mitigate such influences, we calibrate the open-sourced 16 reports to align with the updated \verb|MITRE ATT&CK|.
We hired two staffs who have at least two years of experience in SOCs for technique labeling. Disagreements were resolved by expert review. This inter-annotator agreement was high (\textbf{Cahen's Kappa=0.89}). 

We additionally collected 1,769 CTI reports from Cisco as of October 2024~\cite{cisco-CTI-report}, categorizing them into medium (1,398) and severe (371) based on CVSS scores. Medium reports fall within the [4.0–8.9] range, while severe reports range from [9.0–10.0], reflecting the varying threat levels encountered in real-world scenarios. 
Most reports do not provide unified and formatted threat intelligence to validate our extracted results, which is also our key motivation behind this work.
We randomly select several techniques extracted from reports for manual investigation to evaluate their accuracy and reliability.

\noindent\textbf{Metrics for Technique Identification.} 
We calculate the false negatives, false positives, precision, recall, F1 scores. We apply the \textbf{exact match criteria} for calculating the metrics:   
\begin{itemize}[fullwidth,itemindent=0em]
\item \textbf{False Negatives (FNs) and False Positives (FPs):} 
With the ground truth of labeled techniques, false negatives refer to the number of correct techniques that are not identified. False positives refer to the number of identified techniques that are not correct. 
True Positives (TPs) refer to the number of recognized techniques that are correct. 
\item \textbf{Precision:} the percentage of correct techniques out of all the identified techniques, represented as $ \frac{TP}{TP+FP}$.
 \item \textbf{Recall:} the percentage of all correctly identified techniques out of all the correct techniques, represented as $\frac{TP}{TP+FN}$.
\item \textbf{F1:} the harmonic mean of precision and recall, represented as $F_1 = 2 \cdot \frac{\text{Precision} \cdot \text{Recall}}{\text{Precision} + \text{Recall}}$.
\end{itemize}

\noindent\textbf{Metrics for Procedure Generation.}
We evaluate the semantic quality for generated procedures from five perspectives below. 
\begin{itemize}[fullwidth,itemindent=0em] 
    \item \textbf{Completeness}: whether these procedures follow the logical structure of <entities relationships actions>.  
    \item \textbf{Correctness}: whether the procedures are technical errors or security misleading.  
    \item \textbf{Relevance}: whether the procedures reflect a real-world or simulated attack scenario. 
    \item \textbf{Reproducibility}: 
    can the procedures be reproduced precisely?  
    \item \textbf{Variativeness:} whether the generated procedures represent plausible procedure variants?
\end{itemize}

\subsection{Down-stream Tasks}  

\subsubsection{Sigma Rule Generator} 
To simulate real-world incident response behaviors, we assume that an expert summarizes CTI reports based on attacks observed over a specific period, and these reports are then analyzed for further incident response activities in the subsequent months. To this end, we deployed a honeypot to safely attract and collect real-world application event logs, as rule-based detection is particularly effective for identifying individual/isolated malicious events that are more frequently captured in application logs (e.g., SQL injection, unauthorized access, file uploads).  
We collect logs over a three-month period from July 22 to October 22, 2024, comprising 64,185 security events. Our experienced analysts summarized CTI reports based on logs collected during the first month, resulting in 32 reports, to test the logs from the following two months. 


\noindent We compare the following Sigma rules-set in our experiments.  
\begin{itemize}[fullwidth,itemindent=0em]
    \item \textbf{\name}: Our approach uses ATI-guidance and environment feedback. 
    \item \textbf{LLM-generated rules}: Standalone LLMs without extracted ATIs or environmental feedback.
    \item \textbf{Open-source rules:} Rules from public repositories such as Splunk~\cite{Splunk-rules} and Sigma~\cite{Sigma}, filtered for web application attacks and comparable rule count.
\end{itemize}

\noindent\textbf{Implementation Details.}
We evaluate detection performance using Splunk version 9.3.1 with a trial license. Logs from August 20 to October 22 were ingested and analyzed by each rule-set.  
We utilize a Cowrie honeypot~\cite{honeypot} deployed on an Amazon Web Services (AWS) server under the three-month free plan.
The Cowrie honeypot records the web attacks~\cite{DBLP:conf/eurosp/NawrockiKHKSW23, DBLP:conf/ndss/WangHLJYZRCCGC20, DBLP:conf/sp/RossowDGKPPBS12, pauley2023understanding} in the wild including brute-force attacks (e.g., brute force login via SSH or Telnet to try different combinations.), command injection (e.g., the vulnerable commands), file operation (e.g., modify files to create or alter users, maintaining persistent access to the system) or ransomware (e.g., upload and execute ransomware or encrypt files on the system) attacks, etc.
Note that the logs collected by Honeypot are involved with system normal behaviors, as exemplified below, explaining the false positives. Two staffs manually labeled normal security events with Cahen's Kappa of 0.84.  

\begin{itemize}\nonumber
\footnotesize
    \item \textit{2024-07-22,18:19:55,/,``Mozilla/5.0 (X11; Linux x86\_64) AppleWebKit/537.36 (KHTML, like Gecko) Chrome/81.0.4044.129 Safari/537.36'', 185.254.196.173.}
\end{itemize}

\noindent\textbf{Metrics.} 
We ingest the test logs into Splunk, execute rules-set as search queries, and verify their accuracy by comparing retrieved logs against the ground truth. 
Execution is considered successful if the retrieved logs match the expected attack-related malicious logs. 
We test all the rules-set with the same condition, specifically, we regard the malicious events as those originating from the same labeled malicious IP sources in ground truths.


\subsubsection{Command Generator} 
A command is generated with a well-structured procedure. As a baseline, commands are generated from the same CTI reports without procedural structuring and the limit of command numbers. 

\noindent\textbf{Metrics.}
We use the \colorbox{mygray}{\textit{subprocess}} module in Python, which runs system commands to test the executability of the command.  
To ensure a fair comparison, we regard a command as executable if it contains no syntax errors (e.g., ``syntax error near unexpected token''). Other failures caused by system-specific or environment-dependent factors, such as missing directories, are excluded from our evaluation, eliminating environmental biases.
To achieve this, our function calculates the percentage of exit codes, where a \textit{returncode} of 0 indicates none syntax errors. Our evaluation focuses on syntactic validity, as achieving semantic correctness would necessitate subjective methods—such as user studies—which involve high-level behavioral interpretation or integration with threat modeling frameworks. Syntactic validity aligns closely with real-world adversarial behavior, since malicious commands must be syntactically correct in order to execute successfully and achieve their intended effects.


\section{Evaluation~\label{sec:evaluation}} 
In this section, we are keen to investigate the following research questions.  

\begin{itemize}[fullwidth,itemindent=0em]
    \item \textbf{RQ1 Accuracy and Depth on ATI Extraction:} How accurately does \name identify techniques, and how reliable and useful are the generated procedures/variants from five perspectives?    
    \item \textbf{RQ2 Ablation Study:}  How does each component (e.g., \emph{knowledge enhancer} and \emph{validator}) enhance the overall performance?  
    \item \textbf{RQ3 Effectiveness on Down-Stream Tasks:} How does the detection performance of our Sigma rule generators? What is the execution rate of our generated commands? 
    \item \textbf{RQ4 Efficiency:} What is the time latency and economic costs of \name?
  \end{itemize}

\subsection{RQ1: Accuracy and Depth on ATI extraction}

\begin{table}[t]
\centering
\setlength{\abovecaptionskip}{0pt}
\setlength{\belowcaptionskip}{0pt}
\caption{Accuracy of technique identification. GPT-4o-mini (mini); GPT-4o (4o); DeepSeek-V3 (DS); LLaMa-3 (LMa).} 
\label{tab:technique-identification}
\renewcommand\tabcolsep{1pt}
\small
\scalebox{0.9}{
\begin{tabular}{cl|c|c|c|c|cl|c|c|c|c|c}
\toprule[\thickline]
\multirow{3}{*}{\begin{tabular}[c]{@{}c@{}}CTI\\ Reports\end{tabular}} &  & \multicolumn{5}{c}{False Negatives ($\downarrow$)}                         &  & \multicolumn{5}{c}{False Positives ($\downarrow$)}                         \\ \cmidrule{3-7} \cmidrule{9-13} 
&  & \multirow{2}{*}{AttacKG} & \multicolumn{4}{c}{\name}   &  & \multirow{2}{*}{AttacKG} & \multicolumn{4}{c}{\name}   \\ \cmidrule{4-7} \cmidrule{10-13} 
&  &                           & mini & 4o & DS & LMa &  &                           & mini & 4o & DS & LMa \\ \cmidrule{1-1} \cmidrule{3-7} \cmidrule{9-13} 
\multicolumn{1}{l}{\texttt{TC\_Firefox DNS}}                                                     &  & 2                          & \cellcolor{deepgray} 2      & 3   &    4      &    6   &  &     14                      & \cellcolor{deepgray} 2       &  4  &    3      &   4    \\ \cmidrule{1-1} \cmidrule{3-7} \cmidrule{9-13} 
\multicolumn{1}{l}{\texttt{TC\_Firefox Drakon}}                                                  &  &  3                         & \cellcolor{deepgray} 1       & 2   &     3     &\cellcolor{deepgray}  1     &  &    4                       &    2     &   4 &\cellcolor{deepgray}   1       &   3    \\ \cmidrule{1-1} \cmidrule{3-7} \cmidrule{9-13} 
\multicolumn{1}{l}{\texttt{TC\_Firefox BITX}}                                                     &  &  5                         &\cellcolor{deepgray} 2        & 3   &     4     &    6   &  &    7                       &    3     & \cellcolor{deepgray} 2  &  3        &   3    \\ \cmidrule{1-1} \cmidrule{3-7} \cmidrule{9-13} 
\multicolumn{1}{l}{\texttt{TC\_SSH}}                                                     &  &     5                      & \cellcolor{deepgray} 2       &  \cellcolor{deepgray}2   &    3      &   4    &  &         3                  &    \cellcolor{deepgray} 1    &  3  &    3      &     3  \\ \cmidrule{1-1} \cmidrule{3-7} \cmidrule{9-13} 
\multicolumn{1}{l}{\texttt{TC\_Nginx}}                                                     &  &   5                    & \cellcolor{deepgray}  4      &  5  &    5      & \cellcolor{deepgray}   4   &  &      10                     &  \cellcolor{deepgray}1       & 2   &   4       &   4      \\ \cmidrule{1-1} \cmidrule{3-7} \cmidrule{9-13} 
\multicolumn{1}{l}{\texttt{Frankenstein}}                                                &  &     9                   & \cellcolor{deepgray}  3      & \cellcolor{deepgray}3  &     4     &   6    &  &     5                      &  \cellcolor{deepgray} 2      &  4  &   4       &   3    \\ \cmidrule{1-1} \cmidrule{3-7} \cmidrule{9-13} 
\multicolumn{1}{l}{\texttt{OceanLotus}}                                                  &  &  5                         &3        &  \cellcolor{deepgray}1   &  4        &   2    &  & 2                          &  \cellcolor{deepgray} 0      &   1 &    1      &   2    \\ \cmidrule{1-1} \cmidrule{3-7} \cmidrule{9-13} 
\multicolumn{1}{l}{\texttt{Cobalt}}                                                     &  &     10                      &   3     & \cellcolor{deepgray}2  &    4      &   5    &  &         3                  &   \cellcolor{deepgray}  2    & \cellcolor{deepgray} 2  &   3       & \cellcolor{deepgray}  2    \\ \cmidrule{1-1} \cmidrule{3-7} \cmidrule{9-13} 
\multicolumn{1}{l}{\texttt{DeputyDog}}                                               &  &        10                   &  \cellcolor{deepgray}2      &   4 &  \cellcolor{deepgray}    2    &   5    &  &          6                 & \cellcolor{deepgray}  1      &  9  &    5      &  5     \\ \cmidrule{1-1} \cmidrule{3-7} \cmidrule{9-13} 
\multicolumn{1}{l}{\texttt{HawkEye}}                                                     &  &    19                     &  \cellcolor{deepgray} 2      &  4  &      9    &   8    &  &          8                 &    3     & \cellcolor{deepgray} 1  &    6      &   6    \\ \cmidrule{1-1} \cmidrule{3-7} \cmidrule{9-13} 
\multicolumn{1}{l}{\texttt{DustySky}}                                                     &  &    5                       &  2      & \cellcolor{deepgray}1  &    3      &   3    &  &         7                  & \cellcolor{deepgray} 3       & 5   &   4       &   6    \\ \cmidrule{1-1} \cmidrule{3-7} \cmidrule{9-13} 
\multicolumn{1}{l}{\texttt{TrickLoad Spyware}}                                          &  &     7                      & 2        & \cellcolor{deepgray}1   &   4       &   2    &  &        3                   &  \cellcolor{deepgray}  1     &  3  &   6       & 6      \\ \cmidrule{1-1} \cmidrule{3-7} \cmidrule{9-13} 
\multicolumn{1}{l}{\texttt{Emotet}}                                                     &  &    5                       &  3       &  \cellcolor{deepgray}1  &     4     &  2     &  &       4                    &  \cellcolor{deepgray} 2      &   5 &   4       &   5    \\ \cmidrule{1-1} \cmidrule{3-7} \cmidrule{9-13} 
\multicolumn{1}{l}{\texttt{Uroburos}}                                                 &  &   7                        & \cellcolor{deepgray}  3      &  4  &   \cellcolor{deepgray}  3     &   5    &  &         5                  &  3       &   5 &   \cellcolor{deepgray} 2      &    \cellcolor{deepgray}2   \\ \cmidrule{1-1} \cmidrule{3-7} \cmidrule{9-13} 
\multicolumn{1}{l}{\texttt{APT41}}                                                     &  &   9                        &   4      &  \cellcolor{deepgray}3  &     7     &    4   &  &         11                  & \cellcolor{deepgray} 3       & 6   &     6     &  7     \\ \cmidrule{1-1} \cmidrule{3-7} \cmidrule{9-13} 
\multicolumn{1}{l}{\texttt{Espionage}}                                                  &  &  10                         &  \cellcolor{deepgray} 3      & 4   &   5       &  3     &  &         3                  & \cellcolor{deepgray} 1       &  2  &    4      &   13    \\ \bottomrule
\multicolumn{1}{l}{Precision $\uparrow$}                                                     &  &      0.313                     & \cellcolor{deepgray}  0.818      &  0.696  &   0.653       &  0.654     & &              --             &   --      &  --  &      --    & --      \\ \cmidrule{1-1} \cmidrule{3-7} \cmidrule{9-13} 
\multicolumn{1}{l}{Recall $\uparrow$}                                                     &  &        0.366                   & \cellcolor{deepgray}   0.767     &  0.756  &   0.614       &  0.625     &  &  --                         &    --     &  --  &     --     &    --   \\ \cmidrule{1-1} \cmidrule{3-7} \cmidrule{9-13} 
\multicolumn{1}{l}{F1 score $\uparrow$}                                                     &  &     0.338                      & \cellcolor{deepgray}  0.792      &  0.724  &  0.629        &   0.631    &  &   --                        &  --       & --   &    --      & --      \\  \bottomrule[\thickline]

\end{tabular}
}
\end{table}

\begin{figure}[h]
\setlength{\abovecaptionskip}{0pt}
\setlength{\belowcaptionskip}{0pt}
\centering
\scalebox{0.55}{
\includegraphics[width=\linewidth]{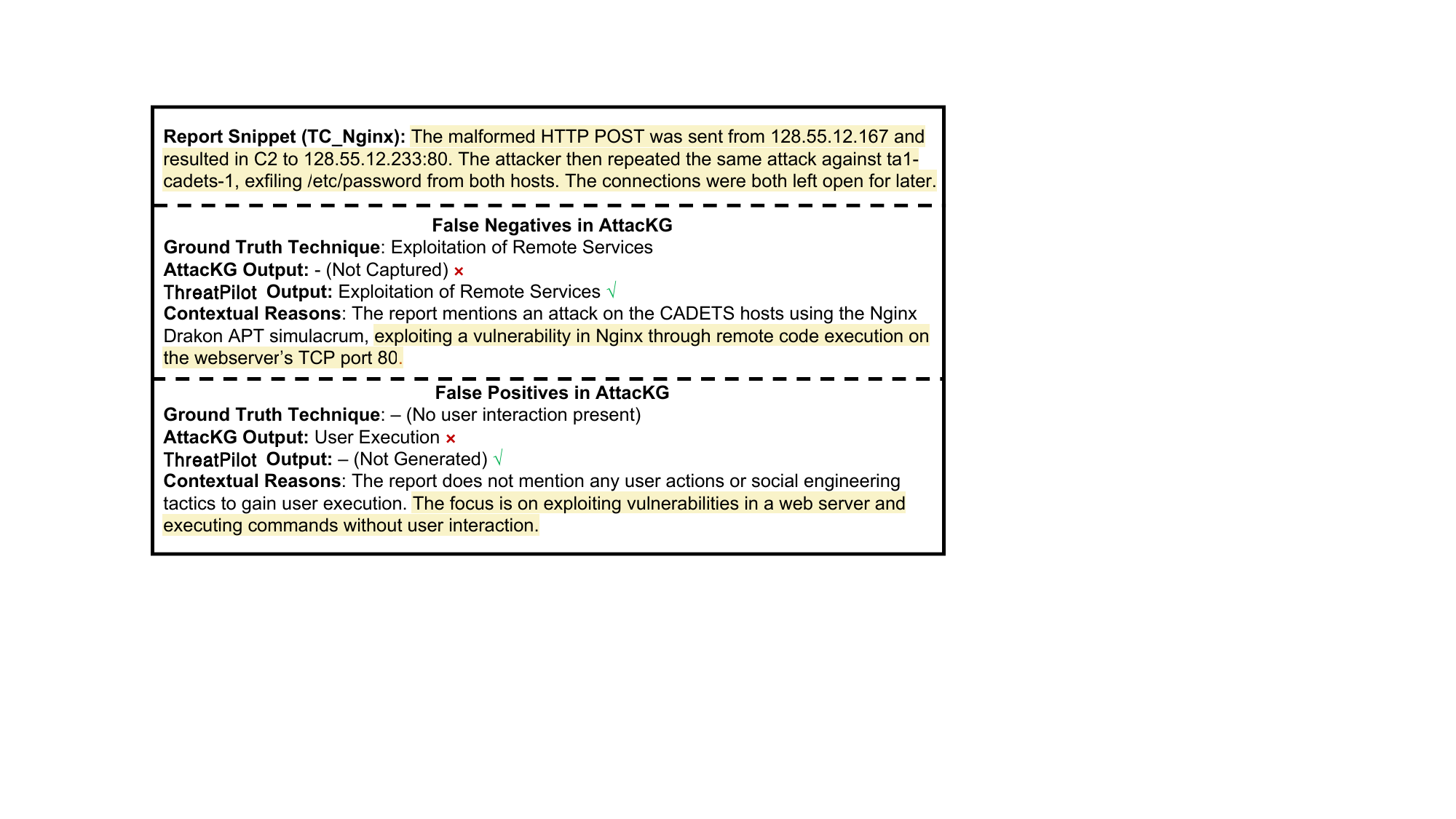}} 
\caption{Examples misflagged by AttacKG but correctly classified by \name.}
\label{fig:result-example}
\end{figure}

\noindent\textbf{Results on Open-sourced CTI Reports.}
Table~\ref{tab:technique-identification} shows the accuracy of ATI extraction on open-sourced CTI reports. \name are more accurate in identifying techniques than AttacKG~\cite{DBLP:conf/esorics/LiZCL22:attackKG} in precision, recall and F1, and upon closed-sourced or open-sourced models. 
To further evaluate \name’s capability in technique extraction, we manually examine the false positives and false negatives that were missed by AttacKG but successfully identified by \name, as illustrated in Figure~\ref{fig:result-example}.
In the \texttt{TC\_Nginx} report, AttacKG fails to identify the technique of \texttt{Exploitation of Remote Services} due to its inability to parse the nounced semantics of remote code execution on the web server’s TCP port 80, which aligns with the reports' description. 
Regarding false positives, 
The \texttt{TC\_Nginx} is unrelated to any user-driven actions, because AttacKG cannot distinguish the fuzzy semantics regarding execution classification.

We attribute this to the fact that \name is better at understanding nuanced attack intentions and comprehending complex natural language expressions including ambiguous descriptions and context-dependent information. In contrast, AttacKG's semantic understanding ability is constrained by its training data and reliance on specific keywords and fixed patterns, making it less effective at parsing nuanced semantics. 

\noindent\textbf{Results on Recently-collected CTI Reports.}
Table~\ref{tab:top-10-techniques} lists the five most common techniques and their frequency based on the recent CTI reports. 
We observe that the distribution of employed techniques in the CTI reports aligns with existing TTP trends~\cite{top-10-techniques,Cyber-proof,picus:top-10}. For example, the technique \texttt{Command and Scripting Interpreter}, which highlights the use of malicious scripts, is widely adopted across a variety of attacks.
Moreover, the top four techniques show significant overlap between medium and severe reports, reinforcing the accuracy of \name, as it reflects the realistic scenario where attackers, especially those from similar threat groups, often reuse a common set of techniques.
In general, the most frequently occurring techniques tend to remain consistent across different CTI reports, with only minor variations.

\begin{table}[t]
\setlength{\abovecaptionskip}{0pt}
\setlength{\belowcaptionskip}{0pt}
\caption{Top five techniques identified from recent-crawled CTI reports. M $\Rightarrow$ Medium report, S $\Rightarrow$ Severe report.} 
\label{tab:top-10-techniques}
\renewcommand\tabcolsep{25pt}    
\centering
\scalebox{0.9}{
\begin{tabular}{c|c|c}
\toprule[\thickline]
Rank & Technique                              & Count \\ \midrule
1    & \texttt{T-1203: Exploitation for Client Execution} (M, S)      & 940 (M), 318 (S) \\ \midrule
2    & \texttt{T-1068: Exploitation for Privilege Escalation} (M, S)  & 561 (M), 200 (S) \\ \midrule
3    & \texttt{T-1059: Command and Scripting Interpreter} (M, S)      & 474 (M), 186 (S) \\ \midrule
4    & \texttt{T-1190: Exploit Public-Facing Application} (M, S)      & 371 (M), 164 (S) \\ \midrule
5  & \begin{tabular}[c]{@{}l@{}}\texttt{T-1055: Process Injection} (M)  \\ \texttt{T-1548: Abuse Elevation Control Mechanism} (S) \end{tabular}
     & 299 (M), 78 (S) \\ \bottomrule[\thickline]
\end{tabular}} 
\end{table}

\noindent\textbf{LLM's Comparison.} 
The closed-source foundation models often outperform open-source ones, primarily due to they are pre-trained with larger parameters. 
Among them, GPT-4o-mini performs best.
Furthermore, both GPT-4o and GPT-4o-mini show similar level of false negatives, but GPT-4o generates significantly more false positives. 
\textbf{We recommend using GPT-4o-mini in \name.}
We speculate that this is due to GPT-4o's creativity to produce more divergent responses, leading to hallucinations and incorrect answers. Given that GPT-4o is around 20x more expensive than GPT-4o-mini while offering minimal additional gains. 

\noindent\textbf{Results on Procedure Quality.}   
We qualitatively evaluate the procedures generated by \name and that by standalone LLMs (i.e., GPT-4o-mini) without the GraphRAG-based reasoning module.  
We randomly sample 10\% recent severe reports to generate procedures, yielding 1,132 and 314 procedures by \name and baseline respectively. 
We revise our prompts in LLM-evaluator to achieve alignment with human evaluators (Cohen’s Kappa = 0.85). 

\begin{figure}[t]
\centering

\subfigure[Quality ratings between the procedures generated by \name and the standalone GPT-4o-mini.  
Corr $\rightarrow$ Correctness, Comp $\rightarrow$ Completeness,  
Rele $\rightarrow$ Relevance, Repr $\rightarrow$ Reproducibility,  
Vari $\rightarrow$ Variativeness.]{
    \scalebox{0.4}{
        \includegraphics[width=\linewidth]{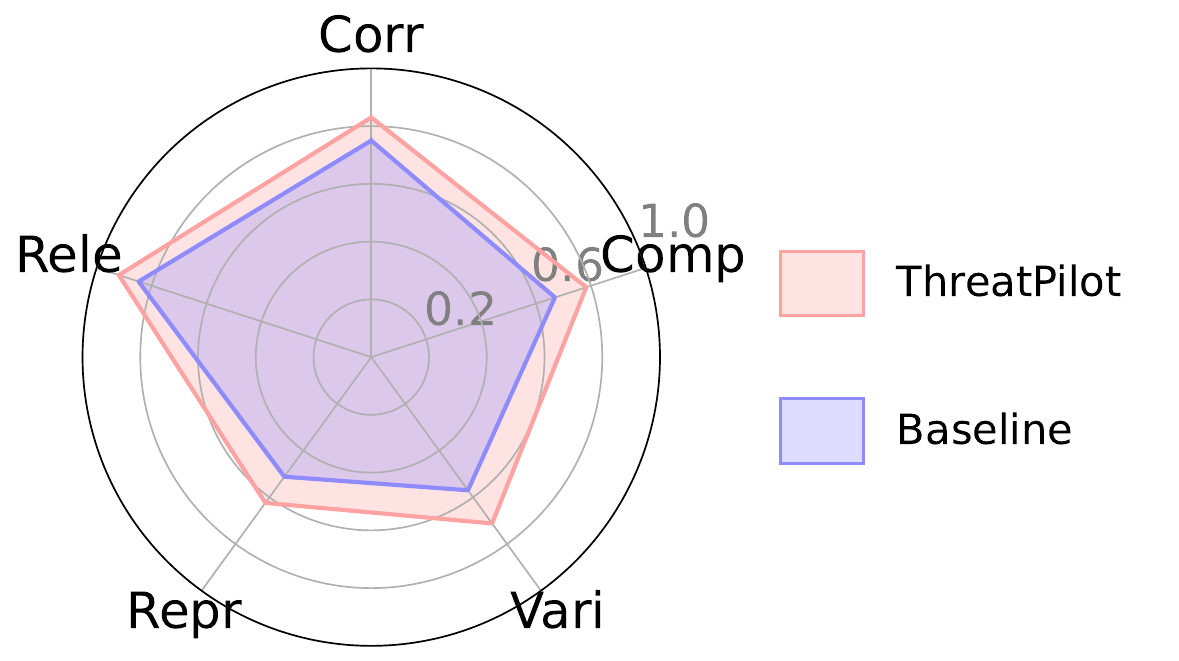}
    }
    \label{fig:procedure-example}
}
\hspace{0.1cm}
\subfigure[Ablation study: with two components, F1 scores can be significantly enhanced.]{
    \scalebox{0.4}{
        \includegraphics[width=\linewidth]{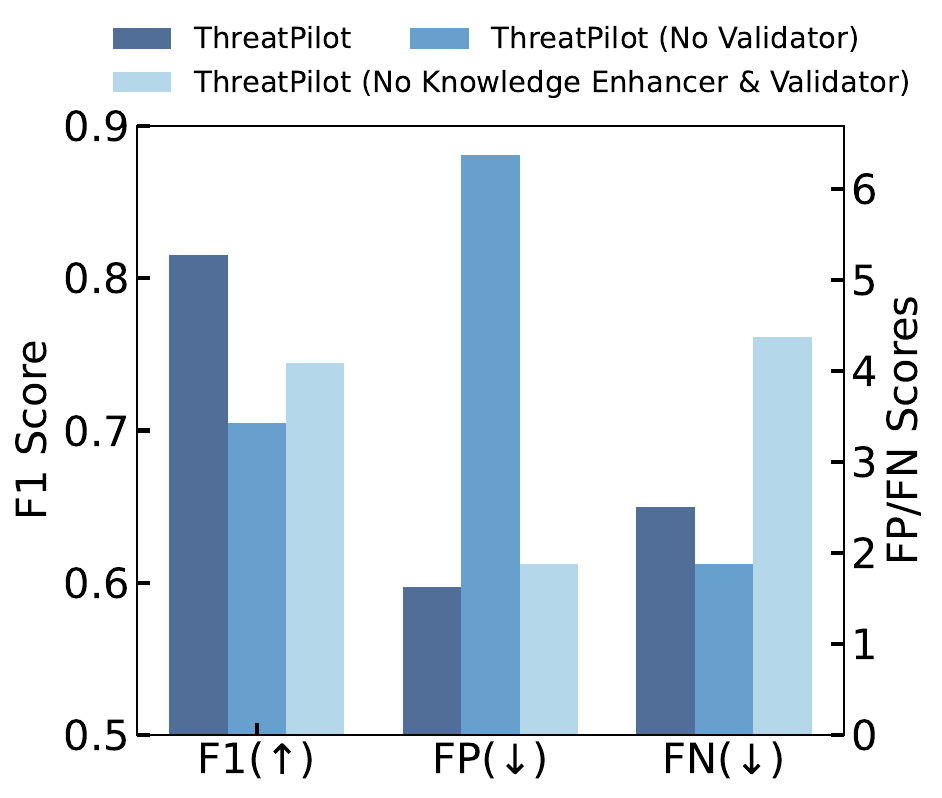}
    }
    \label{fig:ablation}
}

\caption{Comparison of procedure quality and ablation study results.}
\label{fig:combined-procedure-ablation}
\end{figure}

Figure~\ref{fig:procedure-example} presents the average scores per procedure across five evaluation dimensions, demonstrating that \name consistently outperforms the baseline in all aspects. Note that \name infers the procedure variants by the grounded database embedded in a Graph-based database, making it less prone from the LLM's hallucination.
Furthermore, our security experts manually checked that 83.5\% (945/1132) are extremely useful based on their daily experience.  
Take a specific example, the procedures without the GraphRAG-based reasoning components are too general without the entities or relationships, like \texttt{Attacker executes commands using the Command and Scripting Interpreter}.
With \name, the procedures can become more targeted and meaningful, such as \texttt{Attacker manipulates system behavior to gain unauthorized access to files outside of the /ltrx\_user/ directory}.


\subsection{RQ2: Ablation Study}


We conducted an ablation study to evaluate the individual contributions of two critical components of \emph{knowledge enhancer} and \emph{validator}. 
We construct several attack variants: one without both \emph{knowledge enhancer} and \emph{validator}, another without \emph{validator}, and the full version of \name that incorporates both components. 
Our ablation study is based on GPT-4o-mini. 
We present the results in Figure~\ref{fig:ablation}, where the full version of \name achieves the highest F1 score, demonstrating that the combination of the two components leads to improved technique identification performance. The use of the \emph{knowledge enhancer} alone results in lower F1 scores compared to configurations without both \emph{knowledge enhancer} and \emph{validator}, primarily due to the high false positives. Nevertheless, the \emph{knowledge enhancer} plays a crucial role in significantly reducing false negatives. To address the false positive issue introduced by the enhancer, the \emph{validator} module is essential. Together, these two components work synergistically: the \emph{knowledge enhancer} mitigates false negatives, while the \emph{validator} filters out false positives.


\subsection{RQ3: Effectiveness on Down-stream Tasks~\label{sec:application_scenarios}}

\subsubsection{Sigma Rule Generator}

We systematically evaluate and compare the detection abilities in the wild between the rules-set from \name, standalone LLMs and open-source folders. 
Table~\ref{tab:rules-results} presents the results, demonstrating that \name significantly outperforms LLM-generated or the open-sourced rules with all metrics, showcasing great attack adaptability and detection abilities.  Specifically, \name achieves the highest F1 score of 0.839, with a precision of 0.815 and a recall of 0.886 under GPT-4o-mini, outperforming other rule-sets by a margin of 4–18\% across all metrics. Even under less performant LLMs like LLaMa-3, \name still shows notable improvements over both its no-feedback variant and open-source rules. The standalone LLM-generated Sigma rules can sometimes yield better detection results than the existing open-sourced rules-set from Splunk or Sigma, demonstrating that LLMs have great potential in rule generation. 
\name is better than all the standalone LLMs, and through further analyzing, we find that the gap is mainly reflected in the ability to model semantic intent and contextual adaptability. \name can improve the pertinence and robustness of Sigma rules with complete built-in functionalities and functionally-complete logic, making it effective to detect more targeted and stealthy attacks. Existing Sigma and Splunk rules remain static, detecting generic malicious scripts or files and making them less effective in dynamic malicious activities. 

\begin{table}[t] 
\centering
\setlength{\abovecaptionskip}{0pt}
\setlength{\belowcaptionskip}{0pt}
\caption{Real-world anomaly detection performance across different rule sets. The logs were collected from a honeypot that captured live attacks over a three-month period.
}
\label{tab:rules-results}
\renewcommand\tabcolsep{10.1pt}  
\begin{tabular}{lllllllc}
\toprule[\thickline]
\multicolumn{2}{c}{Rules-set (Size)}                   &  & Precision &  & Recall &  & F1    \\ \cmidrule{1-2} \cmidrule{4-4} \cmidrule{6-6} \cmidrule{8-8} 
\multirow{2}{*}{GPT 4o-mini} & No ATI/Feedback (86)           &  & \cellcolor{gray!30} 0.588     &  &  
 \cellcolor{gray!30} 0.774 &  & \cellcolor{gray!30} 0.668 \\ \cmidrule{2-2} \cmidrule{4-4} \cmidrule{6-6} \cmidrule{8-8} & With ATI (235) &  & \cellcolor{gray!45} 0.787         &  & \cellcolor{gray!45} 0.831       &   & \cellcolor{gray!45}  0.808     \\ \cmidrule{2-2} \cmidrule{4-4} \cmidrule{6-6} \cmidrule{8-8}  
& With AIT/Feedback (235) &  & \cellcolor{gray!65}   0.815       &  &  \cellcolor{gray!65}   0.886    &   & \cellcolor{gray!65} 0.839     \\ \midrule[\thinline]
\multirow{2}{*}{DeepSeek-V3} & No ATI/Feedback (68)           &  & \cellcolor{gray!30}  0.598        &  &   0.471  \cellcolor{gray!30}   &  &  \cellcolor{gray!30} 0.527    \\ \cmidrule{2-2} \cmidrule{4-4} \cmidrule{6-6} \cmidrule{8-8} 
                             & With ATI (249) &  &   
 \cellcolor{gray!45} 0.701       &  &  \cellcolor{gray!45} 0.769       &   &  \cellcolor{gray!45} 0.733     \\ \cmidrule{2-2} \cmidrule{4-4} \cmidrule{6-6} \cmidrule{8-8} & With AIT/Feedback (249) &  & \cellcolor{gray!65} 0.762     &  & \cellcolor{gray!65} 0.831  &  &  
                             \cellcolor{gray!65} 0.795 \\ \midrule[\thinline] 
\multirow{2}{*}{LLaMa-3}     & No ATI/Feedback (73)           &  & \cellcolor{gray!30}  0.614        &  &  \cellcolor{gray!30}  0.297    &  &   0.400 \cellcolor{gray!30}   \\ \cmidrule{2-2} \cmidrule{4-4} \cmidrule{6-6} \cmidrule{8-8}
 & With ATI (274) &  & \cellcolor{gray!45}   0.617       &  &  \cellcolor{gray!45}    0.384    &   & \cellcolor{gray!45}   0.473     \\ \cmidrule{2-2} \cmidrule{4-4} \cmidrule{6-6} \cmidrule{8-8} 
 & With AIT/Feedback (274) &  &  \cellcolor{gray!65} 0.697     &  &  \cellcolor{gray!65} 0.471  &  &  \cellcolor{gray!65} 0.562 \\ \midrule[\thinline]
\multicolumn{2}{l}{Splunk~\cite{Splunk-rules}: Application and web (121)}   &  & 0.449     &  & 0.618  &  & 0.520 \\ \cmidrule{1-2} \cmidrule{4-4} \cmidrule{6-6} \cmidrule{8-8} 
\multicolumn{2}{l}{Sigma~\cite{Sigma}: Emerging threat rules (366)}  &  & 0.350     &  & 0.583  &  & 0.483 \\ \bottomrule[\thickline]
\end{tabular}
\end{table}


\noindent\textbf{Case Study.} 
We conducted a case study analyzing both the malicious activities it uniquely detected and the instances it failed to identify.
For uniquely detected cases, \name successfully uncovered threat-targeted behaviors such as privilege abuse, sensitive file access (e.g., reading/writing to \texttt{ADMIN}), and output redirection to remote paths like \texttt{127.0.0.1\textbackslash\textbackslash ADMIN\$}, which may indicate data exfiltration or reverse shell attempts. Unlike traditional static rules that typically flag generic payloads (e.g., Build.bat), \name precisely identifies threat-critical artifacts such as \texttt{Invoke-Expression}, \texttt{DestinationIp}, and \texttt{.env}, \texttt{.git/config}, \texttt{/phpMyAdmin/index.php} from more empirical observations, due to its deep modeling of attack intent, semantic context, and integration of environmental feedback.
Moreover, \name exhibits robustness to feature-shift attacks by detecting variants that static rules-set often miss, demonstrating its intelligence and adaptability in rule generation. 
In contrast, the undetected cases often involve stealthy attacks with entirely novel signatures or behaviors that differ significantly from known patterns, causing both static matching and semantic reasoning methods to miss them.
Finally, we simulate system attacks using logs labeled with ground-truth malicious commands. On average, the rules generated by \name consistently achieve higher F1 scores.

\begin{figure}[t]
\setlength{\abovecaptionskip}{0pt}
\setlength{\belowcaptionskip}{0pt}
\centering

\subfigure[Comparison of Sigma rules with/without ATIs.]{
    \scalebox{0.4}{
        \includegraphics[width=\linewidth]{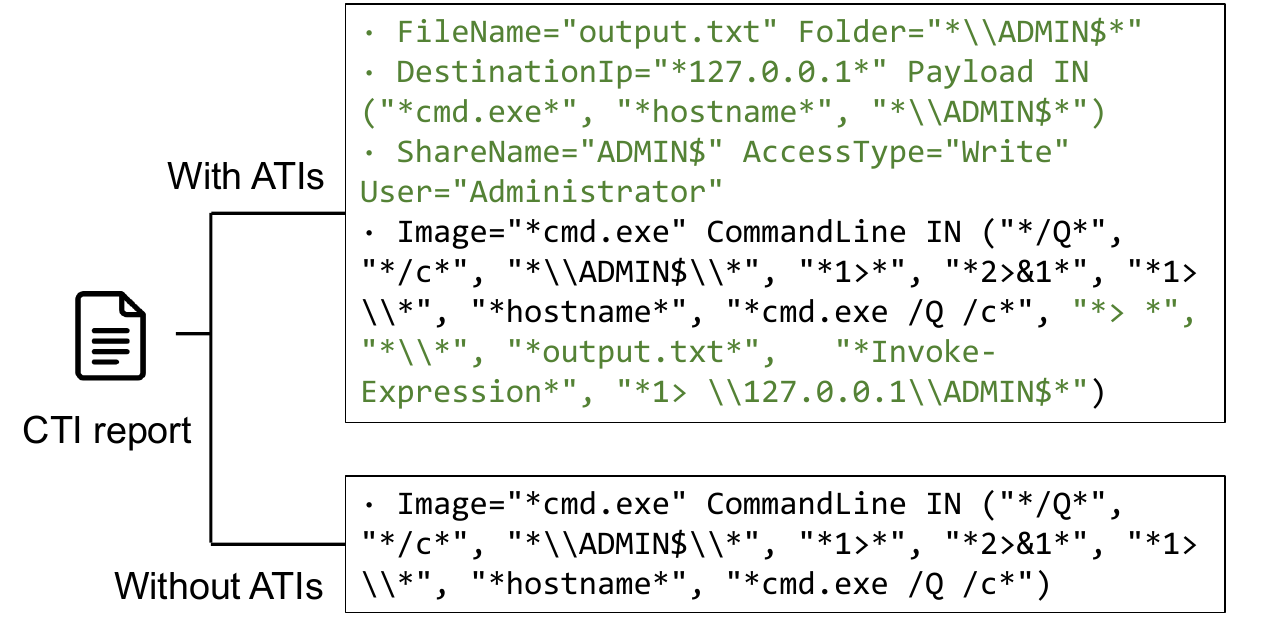}
    }
    \label{fig:Sigma-rule-example}
}
\hspace{0.02\textwidth}
\subfigure[Comparison of commands with/without procedures.]{
    \scalebox{0.54}{
        \includegraphics[width=\linewidth]{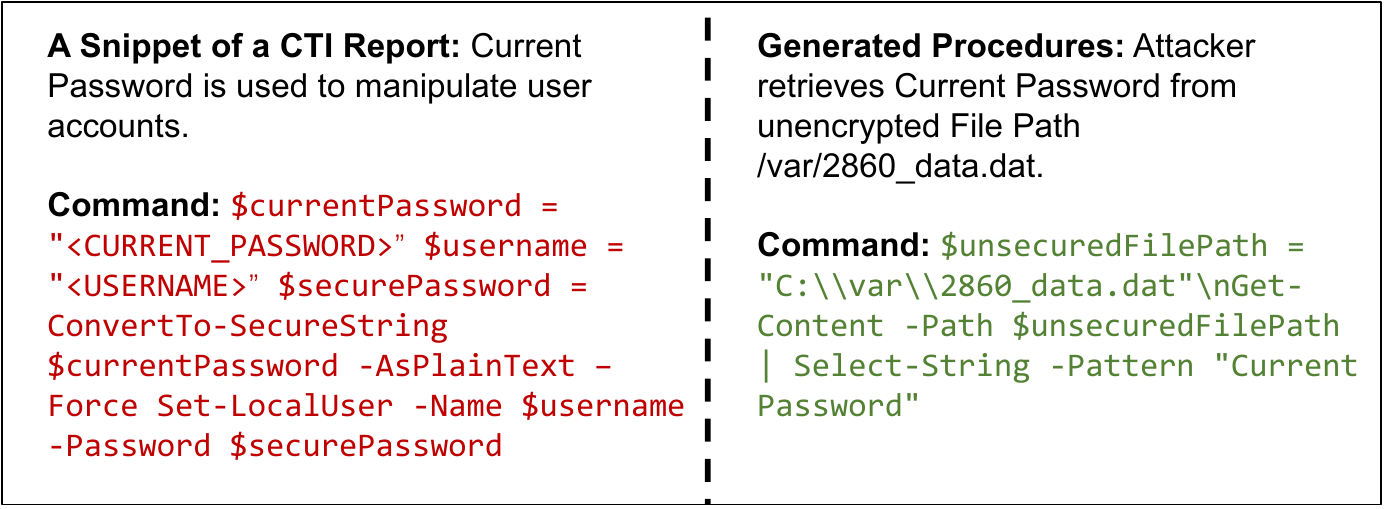}
    }
    \label{fig:command-example}
}

\caption{Comparison of Sigma rules and commands with/without additional information.}
\label{fig:combined}
\end{figure}



\subsubsection{Command Generator} 
We selected the useful 945 higher-rated procedures to generate corresponding commands. 
As shown in Table~\ref{tab:command-useful}, comparing to those without procedures, the commands generated from procedures exhibit a higher executable rate. 
We attribute this to the fact that standardized procedures help mitigate LLM hallucinations, leading to more executable commands, whereas unstructured reports make it challenging for LLMs to comprehend, thereby leading to the failure of execution.
Moreover, Figure~\ref{fig:command-example} presents an example illustrating our interpretation: procedures processed by GraphRAG contain significantly more entities, resulting in more meaningful commands compared to the commands generated by unstructured reports.


\begin{table}[]
\centering
\setlength{\abovecaptionskip}{0pt}
\setlength{\belowcaptionskip}{0pt}
\caption{Execution success of the generated commands.} 
\label{tab:command-useful}
\renewcommand\tabcolsep{21.9pt}
\begin{tabular}{lllc}
\toprule[\thickline]
\multicolumn{2}{c}{Commands (Size)}                          &  & Executable Percentage \\ \cmidrule{1-2} \cmidrule{4-4} 
\multirow{3}{*}{GPT 4o-mini} & No Procedure/Feedback (204) &  & 53.4\%                \\ \cmidrule{2-2} \cmidrule{4-4} 
                             & With Procedure (945)          &  & \cellcolor{gray!40} 86.6\%                \\ \cmidrule{2-2} \cmidrule{4-4} 
                             & With Procedure/Feedback (945) &  & \cellcolor{gray!65} 99.3\%                \\ \midrule[\thinline]
\multirow{3}{*}{DeepSeek-V3} & No Procedure/Feedback (251)      &  &     55.0\%                  \\ \cmidrule{2-2} \cmidrule{4-4} 
                             & With Procedure (879)                &  & \cellcolor{gray!40}  93.4\%                    \\ \cmidrule{2-2} \cmidrule{4-4} 
                             & With Procedure/Feedback (879)       &  &  \cellcolor{gray!65}  96.1\%                    \\ \midrule[\thinline]
\multirow{3}{*}{LLaMa-3}     & No Procedure/Feedback (218)         &  &  50.5\%                     \\ \cmidrule{2-2} \cmidrule{4-4}
                             & With Procedure (804)                &  & \cellcolor{gray!40}  86.4\%                    \\ \cmidrule{2-2} \cmidrule{4-4}
                             & With Procedure/Feedback (804) &  &   \cellcolor{gray!65} 93.4\%                   \\ \bottomrule[\thickline]
\end{tabular}
\end{table}

\subsection{RQ4: Efficiency}
We use CTI reports with an average length of 408 words for the efficiency evaluation, .

\noindent\textbf{Latency for ATI Extractor.} 
We show the latency analysis and the economic costs for identifying techniques of \name in Table~\ref{tab:efficiency}.
\name extracts accurate intelligence from CTI reports within seconds, demonstrating its training-free and lightweight design—well-suited for processing live streams of threat intelligence data.

\begin{minipage}[c]{0.3\textwidth}
\centering
\footnotesize
\renewcommand\tabcolsep{1.7pt}
\begin{tabular}{llllll}
\toprule[\thickline]
Model       &  & \begin{tabular}[c]{@{}l@{}}Prompt\\ Tokens\end{tabular} & \begin{tabular}[c]{@{}l@{}}Output\\ Tokens\end{tabular}  & \begin{tabular}[c]{@{}l@{}}Cost\\ \end{tabular} & \begin{tabular}[c]{@{}l@{}}Query\\ Time\end{tabular}  \\ \cmidrule{1-1} \cmidrule{3-6} 
GPT-4o-mini &  & 96,481  & 2,533      & \$0.020                                                  & 56s  \\ \cmidrule{1-1} \cmidrule{3-6} 
GPT-4o      &  &   91,642  & 2,068         &     \$0.488                                                    &  58s      \\
\cmidrule{1-1} \cmidrule{3-6} 
DeepSeek-V3      &  &   173,846  &  4,187       &     --                                                    &  61s     \\
\cmidrule{1-1} \cmidrule{3-6} 
LLaMa-3      &  &   65,313  & 1,584         &     --                                                    &  73s     \\
\bottomrule[\thickline]
\end{tabular}
\captionof{table}{Efficiency of technique identification per CTI report.\label{tab:efficiency}}
\end{minipage}
\begin{minipage}[c]{0.7\textwidth}
\centering
\footnotesize
\begin{tabular}{llccllll}
\toprule[\thickline]
\multicolumn{2}{c}{Tasks}       &  & Ave. Rules/Commands &  & Cost    &  & Time   \\ \cmidrule{1-2} \cmidrule{4-4} \cmidrule{6-6} \cmidrule{8-8} 
\multirow{2}{*}{GPT 4o-mini} & Sigma Rules &  & 18                  &  & \$0.038 &  & 212s \\
\cmidrule{2-2} \cmidrule{4-4} \cmidrule{6-6} \cmidrule{8-8}
                             & Commands    &  & 24                 &  &    \$0.045     &  &    246s    \\ \cmidrule{1-2} \cmidrule{4-4} \cmidrule{6-6} \cmidrule{8-8} 
\multirow{2}{*}{DeepSeek-V3} & Sigma Rules &  & 17                  &  & --      &  & 225s   \\ \cmidrule{2-2} \cmidrule{4-4} \cmidrule{6-6} \cmidrule{8-8} 
                             & Commands    &  & 20                  &  & --      &  & 307s     \\ \cmidrule{1-2} \cmidrule{4-4} \cmidrule{6-6} \cmidrule{8-8} 
\multirow{2}{*}{LLaMa-3}     & Sigma Rules &  & 18                  &  & --      &  & 241s   \\ \cmidrule{2-2} \cmidrule{4-4} \cmidrule{6-6} \cmidrule{8-8} 
                             & Commands    &  & 25                 &  & --      &  & 289s     \\ \bottomrule[\thickline]
\end{tabular}
\captionof{table}{Efficiency of end-to-end generation of Sigma rules/commands per CTI report.\label{tab:end-to-end-efficiency}
}
\end{minipage}

\noindent\textbf{End-to-end Latency Analysis for Sigma Rule Generator.} 
Table~\ref{tab:end-to-end-efficiency} shows an end-to-end efficiency for generating Sigma rules and executable commands from a CTI report. 
\name significantly outperforms manual efforts in terms of both latency and cost-efficiency, as the effort required for producing detection rules is highly dependent on the analyst's expertise and the complexity of the report~\cite{security-money-cost, DBLP:conf/trustcom/RastogiDGZA22:label-CTI}. 

\section{~\label{sec:discuss}Discussion}


\noindent\textbf{Creativity vs. Hallucination.} 
When using LLMs, a key trade-off is balancing creativity and hallucination. Creativity in LLMs allows \name to generate unseen techniques. This capacity is particularly valuable in finding new ways between attack actions. However, the same mechanisms that drive creativity also increase the risk of hallucinations. We manage this trade-off in \name by the combination of \emph{knowledge enhancer} and \emph{validator}. 
Hallucination is a pervasive concern in LLM-assisted work~\cite{mundler2023self,bechard2024reducing}. \name is designed to mitigate this issue while acknowledging its inevitability.
This robustness stems from the dual approach of creative generation followed by a fact-checking process, allowing it suitable for tasks for both creativity and reliability.
Specifically, \name combines a vector database which stores verified information and known techniques along with an \emph{LLM-as-a-judge} component.

\noindent\textbf{Why not Fine-tune LLMs.} 
We opt for LLM augmentation over fine-tuning due to its practical advantages and observed limitations. A preliminary fine-tuning on a dataset of <techniques: descriptions> from \verb|MITRE ATT&CK| yielded suboptimal results in identifying complete techniques. We attribute this to the low quality and limited coverage of the fine-tuning dataset. Furthermore, fine-tuning adjusts all model parameters and thus requires exceptionally high-quality data, which is often hard to obtain in practice.
In contrast, augmentation allows LLMs to retrieve and incorporate external domain knowledge dynamically, making it a lightweight yet effective solution. It is also more resource-efficient: fine-tuning typically demands significant computational power (e.g., GPUs or paid APIs) and incurs high costs. For example, OpenAI charges \$25 per million tokens for fine-tuning GPT-4o. Given the fast-evolving nature of security knowledge, maintaining up-to-date fine-tuned models becomes costly and impractical, making augmentation the more sustainable approach.

\noindent\textbf{LLMs for Threat Intelligence.}
Recent works~\cite{DBLP:journals/corr/abs-2405-04753-AttackKG++, DBLP:journals/corr/abs-2503-03170:AttackSeqBench} have explored the use of LLMs to enhance threat intelligence extraction. However, they often fall short in bridging the gap between extracted intelligence and its practical application. In this paper, we take a first step toward operationalizing threat intelligence by linking it to the generation of Sigma rules and executable commands, thereby demonstrating its practical value in real-world scenarios.

\section{Conclusion}~\label{sec:conclusion}
In this paper, we present \name, a system for extracting and enriching attack-level threat intelligence (ATI) from a CTI report.
The comprehensive ATI can guide the automatic generation of Sigma rules and commands efficiently for attack defense and reproduction, eliminating the need for analysts to manually adhere to strict syntax rules while ensuring sufficient validation. 
We perform a comprehensive evaluation of \name, evaluating the accuracy of extracted techniques and the execution success of generated Sigma rules and commands. \name outperforms state-of-the-art models like AttackKG by 1.34×. Moreover, our generated rules significantly surpass existing rule-sets in detecting real-world malicious events, and our generated commands achieve an execution success rate of 99.3\%, compared to 50.3\% without the intelligence guided. 




\bibliographystyle{ACM-Reference-Format}
\bibliography{longpassword.bib}

\clearpage
\appendix



\subsection{Detailed Prompts and Procedure Generation}~\label{app:detailed-prompts}
In this section, we provide the detailed prompts used in our framework and the details of procedure generations.  
Table~\ref{tab:prompt-detail-IoC}, Table~\ref{tab:prompt-detail-classifying} and Table~\ref{tab:prompt-detail-judgement} present the detailed prompt for all the attack intelligence extraction ranging from the IoC extraction, the specific guidelines used in in-context learning abilities, and the guidelines used in the component of \emph{LLM-as-a-judge}. 
Finally, to tailor the web application logs in real-world log evaluation, we slightly modify the guidelines to precisely customize the generation of Sigma rules and shown the detailed guidelines in Table~\ref{tab:prompt-detail-rule-generation:field-study}.

\begin{table}[h]
\setlength{\abovecaptionskip}{0pt}
\setlength{\belowcaptionskip}{0pt}
\caption{The detailed prompt for IoC extraction.}  
\label{tab:prompt-detail-IoC}
\centering
\small
\begin{tabularx}{\linewidth}{X}

    \toprule
    
    \coloredbox{mygray}{Guidelines}: \\
    1. The IoCs can include IP addresses, domain names, URLs, file hashes, etc.\\
    2. Focus on terms that describe the nature of the attack, the software, or methods being used.\\
    3. The output needs to be in JSON format.\\
    4. If there is no IoC in the report, return an empty list.\\
    The output format is as follows: 
    \{"ioc": ["IOC1", "IOC2", ...]\}\\
    \bottomrule
\end{tabularx}
\end{table} 

\begin{table}[h]
\setlength{\abovecaptionskip}{0pt}
\setlength{\belowcaptionskip}{0pt}
\caption{The detailed guidelines in \emph{intention interpreter}.}          
\label{tab:prompt-detail-classifying}
\centering
\small
\begin{tabularx}{\linewidth}{X}

    \toprule
    
    \coloredbox{mygray}{Guidelines}: \\
    Tactics represent the "why" of an ATT\&CK technique or sub-technique. It is the adversary's tactical goal: the reason for performing an action. For example, an adversary may want to achieve credential access.\\
    Techniques represent 'how' an adversary achieves a tactical goal by performing an action. For example, an adversary may dump credentials to achieve credential access.\\
    - You need to analyse the techniques used in the attack.\\
    - Make sure that the techniques are one of the techniques above.\\
    - The output needs to be in JSON format.\\
    \bottomrule

\end{tabularx}
\end{table}

\begin{table}[h]
\setlength{\abovecaptionskip}{0pt}
\setlength{\belowcaptionskip}{0pt}
\caption{The detailed guidelines in procedure generation.}  
\label{tab:prompt-procedure}
\centering
\small
\begin{tabularx}{\linewidth}{X}
    \toprule
    \coloredbox{mygray}{Guidelines}: \\
    Analyse the implementation details taken by the attacker to achieve their goal.\\
    The procedure usually contains entities, actions, and relationships.\\
    You need to focus on the entities mostly.\\
    For example, the entities can be following:\\
    - IP addresses or domain names such as 168.0.0.1, example.com\\
    - File names or hashes such as ctfhost2.exe\\
    - Programs or software names such as Mimikatz, vssadmin.exe\\
    - And so on\\
    The report is as follows:\\
    \{attack\_report\}\\
    The technique used in the attack is\{technique\}.\\
    \bottomrule
\end{tabularx}
\end{table}

\begin{table}[h]
\caption{The detailed prompt for \emph{LLM-as-a-judge}.}  
\label{tab:prompt-detail-judgement}
\centering
\small
\begin{tabularx}{\linewidth}{X}

    \toprule
    
    
    \coloredbox{mygray}{Guidelines}: \\
    - If the technique exists in the report, you need to output YES and the reason.\\
    - If the technique does not exist in the report, you need to output NO and the reason.\\
    - The output needs to be in JSON format.
    The output format is as follows:\\
    \{"if\_exist": "YES/NO", "reason": "REASON"\}\\
    \bottomrule 
\end{tabularx}
\end{table} 

\begin{table}[t]
\setlength{\abovecaptionskip}{0pt}
\setlength{\belowcaptionskip}{0pt}
\caption{The detailed guidelines for generating Sigma rules towards web applications.}  
\label{tab:prompt-detail-rule-generation:field-study} 
\centering
\small
\begin{tabularx}{\linewidth}{X}
    \toprule
    \coloredbox{mygray}{Guidelines}: 
    The fields in the sigma rule should only be 'user\_agent', 'extracted\_source', 'url', which means the selection should be based on these fields.\\ 
    \bottomrule
\end{tabularx}
\end{table}

\begin{figure}[h]
\footnotesize
\setlength{\abovecaptionskip}{0pt}
\setlength{\belowcaptionskip}{0pt}
  \centering 
    \subfigure[Clusters in GraphRAG's vector database.]{\includegraphics[width=0.23\textwidth]{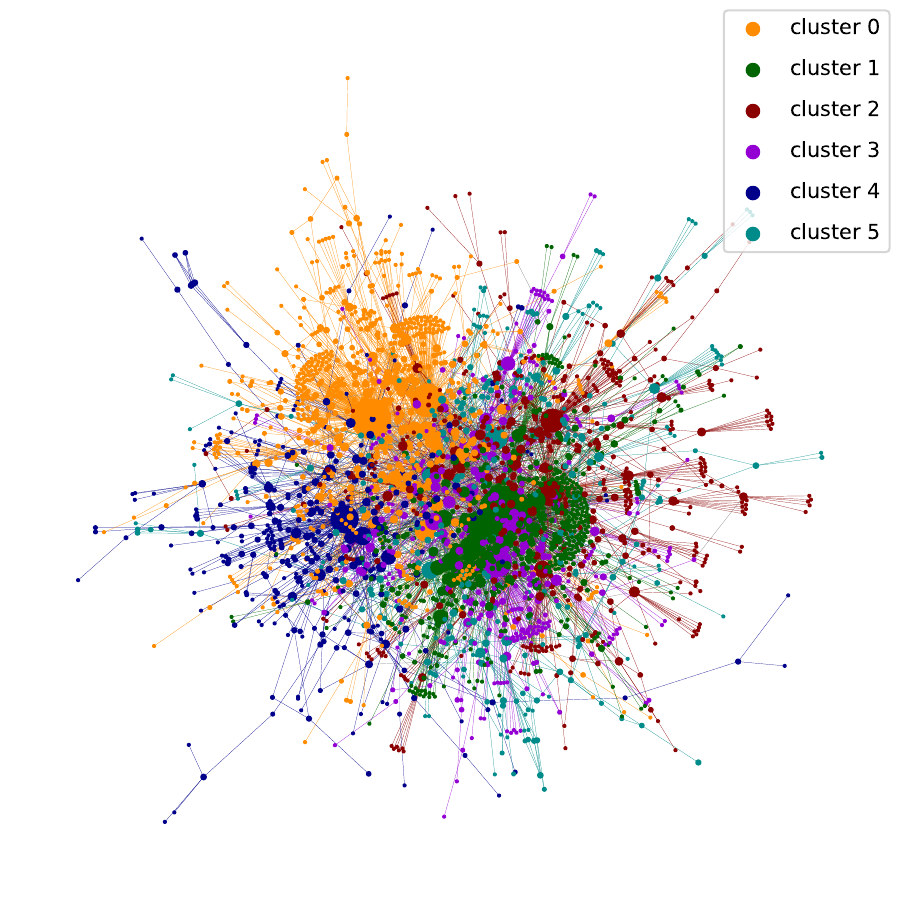}} 
    \subfigure[One example in one cluster.]{\includegraphics[width=0.23\textwidth]{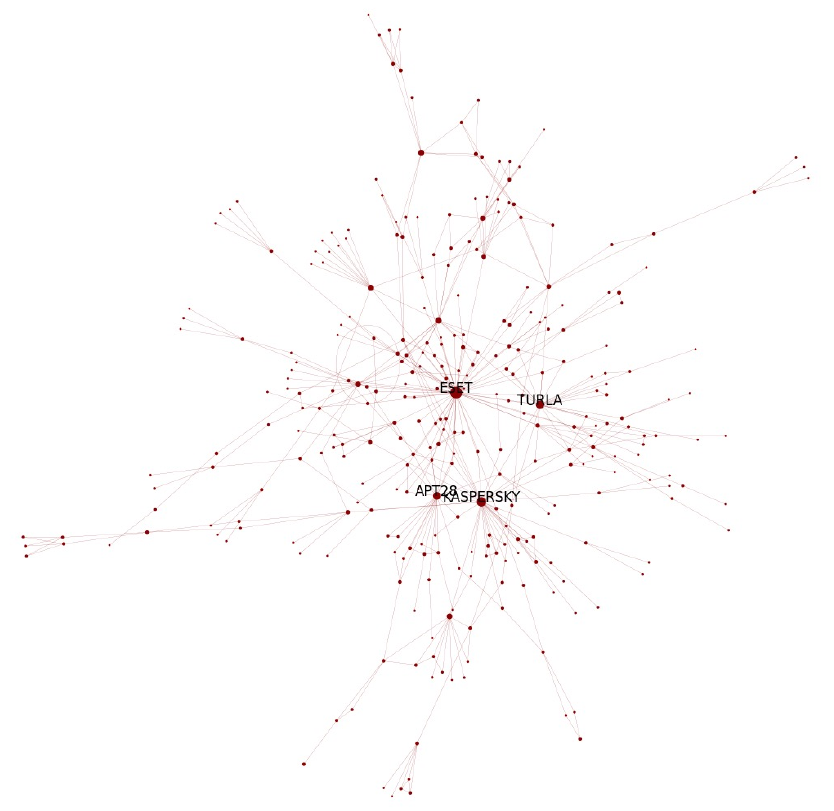}} 
\caption{Demonstration of our vector database in GraphRAG.}  
\label{fig:graphrag-vector-DB}
\end{figure} 

\noindent\textbf{Examples on the vector database in our GraphRAG.} 
Clusters in our vector database of GraphRAG demonstrates the relevant entities and relationships in existing procedures. When generating the new procedures, it could first go through the knowledge graph and then generate the specific procedures. 
Clusters in our vector database of GraphRAG demonstrates the relevant entities and relationships in existing procedures. When generating the new procedures, it could first go through the knowledge graph and then generate the specific procedures.  

\subsection{System Evaluation on Sigma Rule Generator}~\label{app:simulated-datasets}

\noindent\textbf{System Log Collection.} 
To evaluate the quality of our generated Sigma rules, we simulate all the atomic attacks
provided by red teams spanning 12 tactics in MITRE ATT\&CK, which expected to cover all possible attacks in this tactic of attack categories.  
We set a max threshold of 10 for those tactic of attacks with too many atomic tests. 
We summarize our used system logs in Table~\ref{tab:simulated-datasets}  including 229,968 system logs from 61 atomic tests. 
The statistics are listed in Table~\ref{tab:simulated-datasets}, where we cover nearly all tactics of MITRE ATT\&CK. 
The results are listed in Table~\ref{tab:simulated-datasets}.
We only do not simulate one tactic of ``resource development'', because the type is a preparation for attacks including acquiring infrastructure to register domain names for malware, rather than directly executing malicious acts. 
We also open-sourced all the simulated datsets for public use. 
We use EventViewer to collect the system logs including system, sysmon, and PowerShell logs in a virtual machine with the OS of Windows 10 (64-bit). 
Our evaluation focuses primarily on Windows events due to their widespread use in both enterprises and consumer markets. Besides, the sophisticated attacks are mostly designed for Windows platforms. 
Note that our results are suitable for other systems like Linux or FreeBSD.  


\begin{table*}[!t]
\centering
\setlength{\abovecaptionskip}{0pt}
\setlength{\belowcaptionskip}{0pt}
\caption{Summary of our simulated attack logs, comprising 229,968 system logs upon 61 atomic tests. We simulate all tactics in MITRE ATT\&CK, except for resource development due to its largely attack-irrelevant features.} 
\label{tab:simulated-datasets}
\renewcommand\tabcolsep{5pt}
\footnotesize
\scalebox{0.85}{
\begin{tabular}{llllllll}
\toprule[\thickline]
TTP attacks      & \begin{tabular}[c]{@{}l@{}}Reconnaissance\\ T1592.001- \\ Gather Victim\\  Host Information \end{tabular} & \begin{tabular}[c]{@{}l@{}}Resource\\ Development\end{tabular}                 & \begin{tabular}[c]{@{}l@{}}Initial Access\\ T1195-\\ Supply Chain\\  Compromise\end{tabular} & \begin{tabular}[c]{@{}l@{}}Execution\\ T1053.005-\\ Scheduled Task/Job \end{tabular}                                              & \begin{tabular}[c]{@{}l@{}}Persistence \\ T1197-\\ BITS Jobs\end{tabular}                              & \begin{tabular}[c]{@{}l@{}}Privilege Escalation\\ T1055.012-\\ Process Injection \\ Process Hollowing \end{tabular} & \begin{tabular}[c]{@{}l@{}}Defense Evasion\\ T1112-\\ Modify Registry\end{tabular}   \\ \midrule[\thinline]
Atomic Tests     & 1                                                                                                         &   0                                                                                      & 1                                                                                             & 11                                                                                                                                   & 4                                                                                                       & 4                                                                                              & 10                                                                                    \\ \midrule[\thinline]
Number of Logs &  1,438 & 0 & 10,091& 47,269                                                                                                                               & 5,761                                                                                                   & 46,375                                                                                         & 31,560                                                                                \\ \midrule[\thinline]
TTP attacks      & \begin{tabular}[c]{@{}l@{}}Credential Access\\ T1555-\\ Credential from \\ Password Stores\end{tabular}  & \begin{tabular}[c]{@{}l@{}}Discovery\\ T1615-\\ Group Policy \\ Discovery\end{tabular} & \begin{tabular}[c]{@{}l@{}}Lateral Movement\\ T1021.002-\\ Remote Services\end{tabular}      & \begin{tabular}[c]{@{}l@{}}Collection\\ T1557.001-\\ Adversary-in-the-Middle:\\ LLMNR/NBT-NS \\ Poisoning and SMB Relay\end{tabular}  & \begin{tabular}[c]{@{}l@{}}Command and Control\\ T1071.001-\\ Application Layer Protocol
 \\:Web Protocols\end{tabular} &
 \begin{tabular}[c]{@{}l@{}}Exfiltration\\ T1041-\\ Exfiltration Over C2 \\ Channel\end{tabular}                                                                                   & \begin{tabular}[c]{@{}l@{}}Impact\\ T1490-\\ Inhibit System \\ Recovery\end{tabular} \\ 
\midrule[\thinline]
Atomic Tests     & 6                                                                                                         & 5                                                                                       & 4                                                                                             & 1                                                                                                                                    & 3                                                                                                       &   2                                                                                             & 10 \\ \midrule[\thinline]
Number of Logs  & 15,798 & 16,189                                                                                  &  8,752  & 1,910 & 5,530        &  5,954                                                                                               & 33,341 \\ \bottomrule[\thickline]
\end{tabular}
}
\end{table*}

\noindent\textbf{Metrics.} 
We ingest all collected logs into Splunk, execute the rules as search queries on the indexed logs, and verify the accuracy of the generated alarms by comparing them against the ground truth.
For each atomic test associated with a tactic, we gather the corresponding attack reports
from sources such as MITRE ATT\&CK and relevant web corpora. These reports are used to generate rules both with and without TTPs, without restricting the size of the generated rules. On average, each atomic test produces around 10 rules.
For each tactic, we input all the generated rules—amounting to hundreds per tactic—into Splunk's execution window using a combined connector. We then compare all generated alerts with the ground truth associated with that tactic's attacks. Precision, recall, and F1 score are calculated using an exact-match criterion.
Additionally, we measure the number of attempts required to pass the grammar check as another indicator of the quality of the initial rule generation.

\noindent\textbf{Analysis Results.}
With precise TTPs, the effectiveness of the generated rules can be largely improved, in terms of the precision, recall, and attempts to pass the rules' grammar check.
The reasons could be that the detailed TTPs can give a full and organized understanding of the attacks, which can facilitate LLMs to generate better rules. Furthermore, we empirically investigate the separate gains of the module of \emph{LLM-as-a-judge} in our end-to-end rule generation approach. 
We design this module to reject rules generated in previous steps, focusing on reducing false positives. Empirical cases show that the filtered rules are often overly general and attack-irrelevant. For example, a rule like \textit{Image IN (``*CalculatorApp.exe'', ``*hollow\_process\_name.exe'')} is too broad, as it only matches image names without specifically considering the process injection. As a result, the corresponding logs it captures are benign activities, such as loading \textit{CalculatorApp.exe}.

Still, despite our efforts to use precise TTPs to reduce false negatives, critical in industrial operational environments, we are motivated for deep reasons on the remaining false negatives. When we look at several cases, we empirically conclude that the gaps stem from the inherent limitations of Sigma rules. Generating more advanced rules with conditions might address this issue, which we identify as an area for future exploration. 
While Sigma rules are compatible with any SIEM vendor, they primarily rely on single-pattern matching through regular expressions. They lack support for complex queries, such as SQL-style aggregations, which are necessary for statistical analysis or advanced detection. 
For example, when an attacker attempts to exfiltrate sensitive files by downloading multiple `.zip' files from a server, the Sigma rule identifies requests for `.zip' files. However, it is limited to basic string matching and cannot account for contextual factors, such as excessive downloads or filtering benign activities based on user agent strings.

\end{document}